\documentclass[twocolumn,english,prl]{revtex4-1}

\usepackage{amssymb}
\usepackage{amsmath}
\usepackage{graphicx}
\usepackage{amsfonts}
\usepackage[usenames, dvipsnames]{xcolor}
\usepackage{bm}
\usepackage{bbm}
\usepackage{color}
\usepackage{hyperref}
\hypersetup{pdfnewwindow=true,
	colorlinks=true,
	linkcolor=blue,
	citecolor=blue,
	urlcolor=blue}
\usepackage{soul}
\usepackage[caption=false, justification=centerlast]{subfig}
\usepackage{psfrag}
\usepackage{textcomp}
\usepackage{braket}

\usepackage[T1]{fontenc}
\usepackage[latin9]{inputenc}
\usepackage{amsmath}
\usepackage{amssymb}
\usepackage{graphicx}
\usepackage{wasysym}

\makeatletter
\usepackage{braket}
\usepackage{dsfont}

\makeatother

\usepackage{babel}

\begin{document}
\global\long\def\ket#1{\left|#1\right\rangle }%

\global\long\def\bra#1{\left\langle #1\right|}%

\global\long\def\braket#1#2{\left\langle #1\left|#2\right.\right\rangle }%

\global\long\def\ketbra#1#2{\left|#1\right\rangle \left\langle #2\right|}%

\global\long\def\braOket#1#2#3{\left\langle #1\left|#2\right|#3\right\rangle }%

\global\long\def\mc#1{\mathcal{#1}}%

\global\long\def\nrm#1{\left\Vert #1\right\Vert }%

\title{Scalable evaluation of incoherent infidelity in quantum devices }
\author{Jader P. Santos}
\author{I. Henao}
\author{Raam Uzdin}
\email{raam@mail.huji.ac.il}
\affiliation{Fritz Haber Research Center for Molecular Dynamics,Institute of Chemistry,
	The Hebrew University of Jerusalem, Jerusalem 9190401, Israel}

\begin{abstract}
Quantum processors can already execute tasks beyond the reach of classical
simulation, albeit for artificial problems. At this point, it is essential
to design error metrics that test the experimental accuracy of quantum
algorithms with potential for a practical quantum advantage. The distinction
between coherent errors and incoherent errors is crucial, as they
often involve different error suppression tools. The first class encompasses
miscalibrations of control signals and crosstalk, while the latter
is usually related to stochastic events and unwanted interactions
with the environment. We introduce the incoherent infidelity as a
measure of incoherent errors and present a scalable method for measuring
it. This method is applicable to generic quantum evolutions subjected
to time-dependent Markovian noise. Moreover, it provides an error
quantifier for the target circuit, rather than an error averaged over
many circuits or quantum gates. The estimation of the incoherent infidelity
is suitable to assess circuits with sufficiently low error rates,
regardless of the circuit size, which is a natural requirement to
run useful computations. 
\end{abstract}
\maketitle
%\global\long\def\thesection{S-\Roman{section}}
\setcounter{section}{0}
%\global\long\def\thefigure{S\arabic{figure}}
\setcounter{figure}{0}
%\global\long\def\theequation{S\arabic{equation}}
\setcounter{equation}{0}

\onecolumngrid

\section{Introduction}

The premise of quantum technology and quantum computing is to provide
a dramatic improvement over classical devices. In particular, quantum
computers and simulators aim to solve computational problems that
would otherwise require unrealistic resources. Reliability is a key
factor for the success of these devices. External noise and inaccuracies
in the control signals that execute the computation presently limit
the observation of\textcolor{red}{{} }a practical quantum advantage.
Furthermore, the intricate nature of quantum evolution makes error
detection and diagnostics very challenging. As a result of this complexity,
error diagnostic methods such as (linear) cross entropy \cite{arute2019quantum,boixo2018characterizing}
and quantum volume \cite{cross2019validating}, which rely on classical simulations
of the quantum evolution, are not scalable. Experimentally, it is known that although some special states admit a tomographic reconstruction with polynomial resources \cite{cramer2010efficient,lanyon2017efficient}, the tomographic cost is exponential in general \cite{o2016efficient} (even using advanced tomographic techniques like compressed sensing \cite{gross2010quantum}). These experimental and computational hurdles have spurred
the development of more efficient benchmarking approaches \cite{knill2008randomized,magesan2011scalable,magesan2012characterizing,erhard2019characterizing,harper2020efficient,flammia2020efficient}. 

The majority of benchmarking techniques for quantum systems have been
designed to assess the overall performance of a quantum device \cite{buadescu2019quantum,eisert2020quantum,bharti2022noisy}.
This includes randomized benchmarking (RB) and many of its variants
\cite{knill2008randomized,magesan2011scalable,magesan2012characterizing,gaebler2012randomized,wallman2015estimating,wallman2018randomized,proctor2019direct,helsen2019new,chen2022randomized,proctor2022scalable,hines2023demonstrating},
as well as quantum volume \cite{cross2019validating}. In this
holistic approach, which henceforth we will term ``device benchmarking'',
the purpose is to quantify the expected performance of a typical circuit
executed on the benchmarked device. For example,
RB protocols provide a single number that under some conditions is
associated with the average gate infidelity. Device benchmarking methods
can be very useful to study the average behavior of quantum processors
and compare different quantum computing architectures. These methods
are usually based on protocols that involve random quantum circuits,
and therefore they may fail to predict the quality of specific quantum
circuits. In particular, it has been noted that error metrics that
exploit random quantum circuits cannot correctly characterize the
performance of structured quantum algorithms \cite{proctor2022measuring}.

The accurate assessment of quantum circuits is also hindered by the
variety and complexity of errors that take place in quantum computers.
Coherent errors and incoherent errors constitute two broad categories
that occupy a special place in the theory of benchmarking. Coherent
errors represent unitary deviations from the ideal evolution that
a quantum device is instructed to execute. On the other hand, incoherent
(or stochastic) errors arise from the interaction between the system
and the environment, or due to random fluctuations in the control
signals. Coherent errors are known to have a significant impact in
NISQ (noisy intermediate scale quantum) devices \cite{sarovar2020detecting,krinner2020benchmarking,rudinger2021experimental}. Typical
sources for these errors include defective calibration of quantum
gates, or parasitic crosstalk terms like the $Z\otimes Z$ interaction
between superconducting qubits \cite{sung2021realization,kandala2021demonstration}. Yet, the
theoretical framework of benchmarking often relies on the assumption
of stochastic Pauli channels \cite{elben2023randomized}, which describe purely stochastic noise.
While these channels can be engineered via randomized compiling (RC)
\cite{wallman2016noise}, this introduces a potential noise overhead and the resulting
randomized circuit may not reflect the performance of the target circuit
being benchmarked. Moreover, RC operations are applied only on Clifford gates and errors affecting non-Clifford gates are neglected.

In this work, we introduce a method for assessing the quality of a quantum computation, executed by a target quantum circuit, under the assumption of time-dependent Markovian noise. For a given input state, we consider the infidelity between the corresponding output state and its error-free counterpart.
More precisely, our method is useful to measure a contribution to
the infidelity that is exclusively associated with incoherent noise,
which we term ``incoherent infidelity''. Since the evaluation of
this quantity does not involve any randomization of the target evolution,
it provides a faithful measure of the strength of stochastic errors in an arbitrary target circuit.

As compared to device benchmarking, error diagnostics of target circuits
is a less studied topic \cite{magesan2012efficient,onorati2019randomized,proctor2022establishing,carignan2023error,gu2023benchmarking}. In particular, Interleaved RB \cite{magesan2012efficient}
is a RB variant that can be applied to individual Clifford circuits, but it lacks scalability guarantees. Another benchmarking
protocol termed Channel Spectrum Benchmarking \cite{gu2023benchmarking} can in principle
be employed for arbitrary target circuits. However, it requires knowing
the eigendecomposition of the ideal (unitary) circuit to be evaluated,
which may be prohibitive for many interesting cases. Cycle benchmarking
(CB) \cite{erhard2019characterizing} and Cycle Error Reconstruction \cite{carignan2023error} have been
used to characterize shallow portions of a quantum circuit, known
as ``cycles''. More precisely, these methods bechmark ``effective
dressed cycles'', that result from the application of RC to the original
cycle. Later on, an extension of Cycle Error Reconstruction was proposed
for estimating coherent and incoherent error contributions in cycles
\cite{carignan2023estimating}. Although these tools
characterize the performance of single circuits (or cycles), it is
worth mentioning that in all the cases the error metric is an average
over a set of input states. Since this averaging seems to be an integral
part of the underlying protocols, it is also unclear if they can be
adapted to benchmark circuit performance for fixed initial states. 

More recently, a method \cite{proctor2022establishing} using mirror circuits \cite{proctor2022measuring}
was proposed to efficiently estimate the infidelity of arbitrary quantum
circuits in a regime of sufficiently low error rates. This technique
isolates the total circuit infidelity (including coherent and incoherent
errors) from state preparation and measurement (SPAM) errors, and
provides an average of this quantity over initial states. The method
developed here is also robust to SPAM errors of moderate magnitude
and is applicable to circuits with low infidelities. However, it can
extract the incoherent infidelity associated with a fixed initial
state or as an average over a set of input states, thereby providing
additional flexibility to the benchmarking task. Furthermore, it admits
an efficient implementation with computational and experimental resources
that are independent of the size of the system. 

This article is structured as follows. In Sec. \textcolor{blue}{II}, we describe the
basics of quantum mechanics in Liouville space \cite{gyamfi2020fundamentals}, which is
the framework used in our derivations. In Sec. \textcolor{blue}{III}, we characterize
the incoherent infidelity by using a generic master equation that
clearly separates the dynamical contributions of coherent and incoherent
errors. In Sec. \textcolor{blue}{IV}, we introduce our method for estimating the incoherent
infidelity. This method is based on the implementation of circuit
sequences that alternate the target evolution with a proper implementation
of its inverse. In Sec. \textcolor{blue}{V}, we present numerical examples that illustrate
the applicability of our technique in different situations. This involves
circuits subjected to noise, coherent errors, and SPAM errors. In
Sec. \textcolor{blue}{VI}, we provide sufficient conditions for the scalability of our
method, and we present the conclusions in Sec. \textcolor{blue}{VII}. 

\section{Preliminaries and notations}

In the following, $H(t)$ will denote a time-dependent Hamiltonian
and the time derivative will be written as $d_{t}$. For the derivation
of our main results it will be useful to move from Hilbert space to
Liouville space. In Liouville space, a density matrix $\rho$ of dimension
$N\times N$ is flattened into a column ``density vector'' of length
$N^{2}$. This vector can be expressed using a double ket notation, i.e. as $|\rho\rangle\rangle$. As a result, the Liouville von Neumann equation
for unitary dynamics $id_{t}\rho=[H(t),\rho]$ becomes \cite{gyamfi2020fundamentals}:
\begin{align}
id_{t}|\rho\rangle\rangle & =H_{L}(t)|\rho\rangle\rangle,\label{eq:1 SchrodLio}\\
H_{L}(t) & =H(t)\otimes I-I\otimes H(t)^{\textrm{T}},\label{eq:2 HL}
\end{align}
where $I$ is the identity operator in the original Hilbert space
(the $N\times N$ identity matrix) and $H_{L}(t)$ is the Hamiltonian
in Liouville space. Moreover, the superscript $\textrm{T}$ denotes
transposition. 

Since $H_{L}(t)$ is hermitian, the resulting evolution operator in
Liouville space at time $t$, denoted by $U_{L}(t)$, is unitary.
That is, $U_{L}(t)U_{L}^{\dagger}(t)=I_{L}$, where $I_{L}$ is the
$N^{2}\times N^{2}$ identity matrix. Due to the Schrödinger-like
structure of Eq. (\ref{eq:1 SchrodLio}), the solution to $|\rho\rangle\rangle$
takes the simple form $|\rho\rangle\rangle=U_{L}(t)|\rho_{0}\rangle\rangle$, even if
the initial state $\rho_{0}$ is mixed. If the system is open and
the dynamics is Markovian, the Schrödinger-like form \ref{eq:1 SchrodLio}
still holds and reads 
\begin{equation}
d_{t}|\rho\rangle\rangle=\left(-iH_{L}(t)+\mc L(t)\right)|\rho\rangle\rangle,\label{eq:3 H+L SE}
\end{equation}
where $\mc L(t)$ is an operator hereafter called ``dissipator''.
This operator can also be time dependent and characterizes the non-unitary
contribution to the evolution. When the evolution
is described by a Lindblad master equation in Hilbert space, it is
straightforward to obtain $\mc L(t)$. However, the main results of
this paper are independent of the explicit form of $\mc L(t)$. In
particular, we also include dissipators that do not follow the Lindblad
form, and thus are not associated with CPTP maps. Consequently, our
method can quantify infidelity due to probability leakage to an inaccessible
part of the Hilbert space. 

Finally, the standard scalar product between two matrices $A$ and
$B$, given by $\textrm{Tr}[A^{\dagger}B]$ in Hilbert space, reads
$\langle\langle A|B\rangle\rangle$ in Liouville space. Here, $|B\rangle\rangle$ is the Liouville
space vector corresponding to $B$, and $\langle\langle A=|A\rangle\rangle^{\dagger}$.
Accordingly, the expectation value of an operator $A$ takes the form
\begin{equation}
\left\langle A\right\rangle =\textrm{Tr}[A^{\dagger}\rho]=\langle\langle A|\rho\rangle\rangle.\label{eq:4 expect value in Liouville space}
\end{equation}

\section{Incoherent infidelity }

In this section, we introduce and motivate the basic expression for
the incoherent infidelity used in our framework. For simplicity, from
now on we will refer to incoherent errors as ``noise''. The effect
of coherent errors can be characterized by introducing a Hamiltonian
contribution $\delta H_{L}(t)$ to the dynamical equation (\ref{eq:3 H+L SE}).
The resulting equation reads
\begin{align}
d_{t}|\rho\rangle\rangle & =\left(-iH_{L}(t)-i\delta H_{L}(t)+\mc L(t)\right)|\rho\rangle\rangle\nonumber \\
 & =\left(-iH_{L}(t)+\mc L^{\prime}(t)\right)|\rho\rangle\rangle,\label{eq:5 evol including noise and coh errors}
\end{align}
where $\mc L^{\prime}(t)=-i\delta H_{L}(t)+\mc L(t)$. By convention,
the dissipator $\mc L(t)$ represents pure noise in (\ref{eq:5 evol including noise and coh errors}),
and the perfect implementation of the target circuit is generated
by the driving $H_{L}(t)$. Thus, any Hamiltonian deviation from $H_{L}(t)$
is absorbed into $\delta H_{L}(t)$. Note however that we do not need
to specify the form of $\mc L(t)$, for the definition of the incoherent
infidelity or for our derivations. 

Let $T$ be the total evolution time and $\tilde{K}$ the evolution
operator that corresponds to the solution of (\ref{eq:3 H+L SE}),
evaluated at time $T$. This operator represents the noisy implementation
of the target circuit $U_{L}(T)$, in the absence of coherent errors.
Furthermore, let $|\tilde{\varrho}\rangle\rangle=\tilde{K}|\rho_{0}\rangle\rangle$
and $|\varrho^{(\textrm{id})}\rangle\rangle=U_{L}(T)|\rho_{0}\rangle\rangle$ denote
the final states associated with $\tilde{K}$ and $U_{L}(T)$. The
incoherent infidelity quantifies the mismatch between $\varrho^{(\textrm{id})}$
and $\tilde{\varrho}$, and is defined by 
\begin{align}
\varepsilon_{\textrm{inc}}(\varrho^{(\textrm{id})},\tilde{\varrho}) & :=1-F(\varrho^{(\textrm{id})},\tilde{\varrho}),\label{eq:6 incoh infid}\\
F(\varrho^{(\textrm{id})},\tilde{\varrho}) & =\langle\langle \varrho^{(\textrm{id})}|\tilde{\varrho}\rangle\rangle,\label{eq:7 incoh fid}
\end{align}
where $F(\varrho^{(\textrm{id})},\tilde{\varrho})$ is the quantum
fidelity \cite{Nielsen} between $\varrho^{(\textrm{id})}$ and $\tilde{\varrho}$
(using the Liouville space formalism, see Eq. (\ref{eq:4 expect value in Liouville space})).
We remark that Eq. (\ref{eq:7 incoh fid}) is valid for a pure state
$\varrho^{(\textrm{id})}$, which occurs if the initial state $\rho_{0}$
is pure. The incoherent infidelity quantifies the effect of noise in the evolution of the initial state $|\rho_0\rangle\rangle$. Thus, if $\varepsilon_{\textrm{inc}}(\varrho^{(\textrm{id})},\tilde{\varrho})$
is too high for a circuit running an algorithm of interest, it indicates
an excess of noise that should be handled regardless of the calibration
quality of the device. Conversely, error reduction efforts can focus
on calibration if the value of $\varepsilon_{\textrm{inc}}(\varrho^{(\textrm{id})},\tilde{\varrho})$
is sufficiently small. 

To obtain a more explicit expression for $\varepsilon_{\textrm{inc}}(\varrho^{(\textrm{id})},\tilde{\varrho})$,
we solve Eq. (\ref{eq:3 H+L SE}) in the interaction picture and express
the interaction-picture version of $\tilde{K}$ using the Magnus expansion
\cite{blanes2009magnus}. Details on this derivation are provided
in Appendix \textcolor{blue}{I}. We find that 
\begin{equation}
\tilde{K}\approx U_{L}(T)e^{\Omega_{1}},\label{eq:8  K without coh errors}
\end{equation}
where 
\begin{equation}
\Omega_{1}=\int_{0}^{T}\mc L^{int}(t)dt\label{eq:9 Omega1}
\end{equation}
is the first term of the Magnus expansion $\Omega=\sum_{n=1}^{\infty}\Omega_{n}$,
hereafter called ``first Magnus term'', and 

\begin{equation}
\mc L^{int}(t)= U_{L}^{\dagger}(t)\mc L(t)U_{L}(t)\label{eq:10 Lint}
\end{equation}
is the representation of $\mc L(t)$ in the interaction picture. The
symbol $\cong$ is used to indicate that we discard higher-order terms
in the Magnus expansion. For example, $\Omega_{2}$ is given by $\Omega_{2}=\frac{1}{2}\int_{0}^{T}dt'\int_{0}^{t'}[\mc L^{int}(t'),\mc L^{int}(t)]dt.$
We also note that $U_{L}(t)$ is the error-free evolution at time
$t$, obtained from Eq. (\ref{eq:1 SchrodLio}). 

Since $|\tilde{\varrho}\rangle\rangle\cong U_{L}(T)e^{\Omega_{1}}|\rho_{0}\rangle\rangle$,
according to Eq. (\ref{eq:8  K without coh errors}), we can write 

\begin{equation}
F(\varrho^{(\textrm{id})},\tilde{\varrho})\cong\langle\langle \rho_{0}|e^{\Omega_{1}}|\rho_{0}\rangle\rangle.\label{eq:11 approx to incoh fid}
\end{equation}
Next, we consider the regime of weak noise. This regime is characterized by a rescaling of the dissipator of the form  $\mathcal{L}(t)\rightarrow\xi\mathcal{L}(t)$, with a dimensionless parameter $\xi$ such that $\xi\ll1$. Thus, we can 
truncate the exponential $e^{\Omega_{1}}$ and keep the linear-order
approximation $F(\varrho^{(\textrm{id})},\tilde{\varrho})\cong1+\left\langle \Omega_{1}\right\rangle +O\left(\xi^{2}\right)$,
where $\left\langle \cdot\right\rangle :=\langle\rho_{0}|\cdot|\rho_{0}\rangle$.
This yields 
\begin{equation}
\varepsilon_{\textrm{inc}}(\varrho^{(\textrm{id})},\tilde{\varrho})\cong-\left\langle \Omega_{1}\right\rangle +O\left(\xi^{2}\right),\label{eq:12 linear incoh infidelity}
\end{equation}
which is the quantity that we aim to estimate in the following section. 

Finally, we stress that there are various fundamental differences between our theory and the general framework of RB. The First difference is that the linear incoherent infidelity \eqref{eq:12 linear incoh infidelity} constitutes a solid approximation in the regime of weak noise, while RB can handle larger levels of noise. However, under the weak noise condition our method is applicable to target circuits of any size or structure, while RB schemes either benchmark gate sets on average or focus on specific circuits like Clifford gates in the case of Interleaved RB \cite{magesan2012efficient}. In addition, it has been noted that Clifford RB faces scalability issues due to errors accumulated in compiled Clifford gates \cite{proctor2019direct}. In Sec. \textcolor{blue}{VI}, we present scalability conditions that  are potentially easier to meet in practical scenarios. Secondly, RB protocols provide error metrics that average out the initial state, while our theory is designed to evaluate the incoherent infidelity given a fixed initial state. Although the simulation of Sec. \textcolor{blue}{VB} averages the incoherent infidelity over initial states generated by Clifford gates, we do so in order to compare the performance of our method with Interleaved RB, which includes the sampling over initial states as an integral part of its protocol.     

\section{Experimental estimation of the incoherent infidelity }

In practice, the implementation of the target evolution $U_{L}(T)$
includes coherent errors. Moreover, simulating the error-free state
$\varrho^{(\textrm{id})}$ becomes infeasible for sufficiently large
systems. A direct measurement of the incoherent infidelity is hindered
by these obstacles and by state preparation and measurement (SPAM)
errors. Here, we propose a method that approximately provides a SPAM-free
estimation of the quantity $\left\langle \Omega_{1}\right\rangle $,
which can be used to assess the incoherent infidelity $\varepsilon_{\textrm{inc}}(\varrho^{(\textrm{id})},\tilde{\varrho})$
via Eq. (\ref{eq:12 linear incoh infidelity}). 

\subsection{The $K_{I}K$ composite cycle}

Equation (\ref{eq:5 evol including noise and coh errors}) represents
the actual dynamics that takes place during the execution of the target
circuit $U_{L}(T)$. Denoting this imperfect implementation by $K$,
the building block of our method for extracting the incoherent infidelity
is a composite cycle $K_{I}K$, where $K_{I}$ is a specific implementation
of the inverse unitary $U_{L}^{\dagger}(T)$ \cite{henao2023adaptive}. In other words, $K_{I}$ must be such that $K_{I}=U_{L}^{\dagger}(T)$
in the absence of noise and coherent errors. The ideal implementation
of $K_{I}$ and the associated noise are described in Sec. \textcolor{blue}{
IVB}. This results in an extension of the master equation (\ref{eq:5 evol including noise and coh errors})
to the total time interval $(0,2T)$, with $K_{I}$ being implemented
from $t=T$ to $t=2T$. 

An explicit expression for $K_{I}K$ can be obtained by applying the
same strategy that leads to the approximation (\ref{eq:8  K without coh errors}).
Namely, by writing (\ref{eq:5 evol including noise and coh errors})
in the interaction picture and keeping only the first Magnus term
in the corresponding solution. This yields 

\begin{equation}
K_{I}K\cong e^{2\Omega_{1}-i\Theta},\label{eq:13 KIK approx-1}
\end{equation}
where 
\begin{equation}
\Theta=\intop_{0}^{T}U_{L}^{\dagger}(t)\delta H(t)U{}_{L}(t)dt+\intop_{T}^{2T}U_{L}^{\dagger}(2T-t)\delta H(t)U_{L}(2T-t)dt\label{eq:14 accumulated coh error-1}
\end{equation}
is the coherent error accumulated during the cycle $K_{I}K$. More
specifically, the first and second integrals in Eq. (\ref{eq:14 accumulated coh error-1})
account for the coherent errors affecting $K$ and $K_{I}$, respectively. 

To gain some intuition about the use of the composite evolution $K_{I}K$
for measuring the incoherent infidelity, consider first the linear
approximation $e^{2\Omega_{1}-i\Theta}\approx I_{L}+(2\Omega_{1}-i\Theta)$. Similarly to what was done with the dissipator, it is now convenient to introduce a reescaling $-i\delta H_{L}(t)+\mathcal{L}(t)\rightarrow\varsigma\left[-i\delta H_{L}(t)+\mathcal{L}(t)\right]$ that accounts for the strength $\varsigma$ of the total error (coherent and incoherent components). In this way, 
\begin{align}
\langle\rho_{0}|K_{I}K|\rho_{0}\rangle & \cong1+\langle\rho_{0}|\left(2\Omega_{1}-i\Theta\right)|\rho_{0}\rangle+O\left(\varsigma^{2}\right)\nonumber \\
 & =1+2\left\langle \Omega_{1}\right\rangle +O\left(\varsigma^{2}\right),\label{eq:15 linear approx to <KIK>}
\end{align}
where the second line follows from the fact that $\langle\rho_{0}|\Theta|\rho_{0}\rangle=0$.
The proof of this property is given in Appendix \textcolor{blue}{II}.
Equation (\ref{eq:15 linear approx to <KIK>}) is equivalent to the
expression $\left\langle \Omega_{1}\right\rangle \cong\frac{1}{2}\left(\langle\rho_{0}|K_{I}K|\rho_{0}\rangle-1\right)+O\left(\varsigma^{2}\right)$,
which can be used to give a linear estimate of the incoherent infidelity
in Eq. (\ref{eq:12 linear incoh infidelity}). While terms of order $O(\xi^{2})$ are negligible by assumption, $O(\varsigma^{2})$ includes the contribution of the coherent error and thus cannot be directly discarded. To address this issue, in  Sec. \textcolor{blue}{IVC} we
consider certain combinations of circuits that correspond to different
powers of the cycle $K_{I}K$. Given a combination $\sum_{k=0}^{n}a_{k}^{(n)}\left(K_{I}K\right)^{k}$,
which contains $n$ powers of $K_{I}K$, it is possible to estimate
$\left\langle \Omega_{1}\right\rangle $ with an error $O\left(\varsigma^{n+1}\right)$
if the coefficients $a_{k}^{(n)}$ are properly chosen. 

\subsection{Implementation of the inverse evolution $K_{I}$ }

While there are various ways of implementing the inverse of the ideal
circuit, we consider the ``pulse inverse'' $K_{I}$ generated by
the inverse driving \cite{henao2023adaptive}

\begin{equation}
H_{I}(t)=-H(T-t).\label{eq:16 inverse driving}
\end{equation}
The use of $H_{I}(t)$ has a double purpose. In the error-free scenario,
it produces the inverse unitary $U_{L}^{\dagger}(T)$ when applied
during a total time $T$. Hence, we can guarantee that $K_{I}=U_{L}^{\dagger}(T)$
in this case. On the other hand, the use of $H_{I}(t)$ is crucial
for obtaining the operator $2\Omega_{1}$ in the approximation (\ref{eq:13 KIK approx-1}).
Specifically,\textcolor{blue}{{} }the first Magnus term associated with
the pulse inverse $K_{I}$ equals the one corresponding to $K$, and
therefore the noise contribution for the cycle $K_{I}K$ is given
by $2\Omega_{1}$. Such a property allows us to separate the linear
quantity $\left\langle \Omega_{1}\right\rangle $ from higher-order
terms in Eq. (\ref{eq:15 linear approx to <KIK>}). In contrast, other
implementations of $K_{I}K$ produce a noise contribution $\Omega_{1}+\Omega_{I,1}$
to the cycle error, where $\Omega_{I,1}$ stems from $K_{I}$ and
is in general different from $\Omega_{1}$. Since the ensuing residual
term $\left\langle \Omega_{I,1}\right\rangle $ cannot be eliminated
by the method presented in Sec. \textcolor{blue}{IVC}, the pulse inverse $K_{I}$ is
a key element for our estimation of the incoherent infidelity.

Equation (\ref{eq:18 inverse dissipator}) relies on Eq. (\ref{eq:16 inverse driving})
and on the way that the time dependence of $\mathcal{L}(t)$ is affected
by the time dependence of $H(t)$. We assume a quasi-stationary time
dependence, meaning that the noise does not drift over a certain amount
of time $t_{\textrm{max}}$. This condition implies that, for $t\leq t_{\textrm{max}}$,
any cycle involving $H(t)$ is subjected to the same dissipator $\mathcal{L}(t)$.
That is, 
\begin{equation}
\mathcal{L}(t+2mT)=\mathcal{L}(t),\label{eq:17 stationary noise}
\end{equation}
for some positive integer such that $t+2mT\leq t_{\textrm{max}}$.
Here, the time period $2T$ expresses the fact that in circuit sequences
involving the cycle $K_{I}K$ the driving $H(t)$ would be repeated
after $2T$. As explained in Sec. \textcolor{blue}{IVC}, these sequences constitute
the circuits to be executed for our evaluation of the incoherent infidelity.
The total time for the implementation of such circuits must also be
smaller than $t_{\textrm{max}}$, to guarantee the fulfillment of
the condition (\ref{eq:17 stationary noise}). 

The time dependence of $\mathcal{L}(t)$ can be arbitrary so long
as it satisfies Eq. (\ref{eq:17 stationary noise}). In particular,
our approach admits noise mechanisms that can affect different gates
in varying ways. An example of this is leakage noise in systems where
the qubit is encoded in two levels of a larger Hilbert
space, as in superconducting qubits. This leakage noise will depend
in general on the non-adiabatic character of the driving $H(t)$. 

Keeping in mind the irreversible character of noise, we consider that
the instantaneous effect of $\mathcal{L}(t)$ remains unchanged when
the sign of the Hamiltonian is reversed. Thus, $H(t)$ and $-H(t)$
give rise to the same dissipator $\mathcal{L}(t)$. In combination
with Eq. (\ref{eq:16 inverse driving}), this argument leads to the
relationship 

\begin{equation}
\mathcal{L}_{I}(t)=\mathcal{L}(T-t),\label{eq:18 inverse dissipator}
\end{equation}
where $\mathcal{L}_{I}(t)$ is the dissipator that acts alongside
the driving $H_{I}(t)$. The condition of quasi-stationary noise is
also applicable in this case, meaning that $\mathcal{L}_{I}(t+2mT)=\mathcal{L}_{I}(t)$
if $t+2mT\leq t_{\textrm{max}}$. 

To have a clearer picture of Eq. (\ref{eq:18 inverse dissipator}),
consider a discrete set of (possibly time dependent) dissipators $\{\mathcal{L}_{i}\}_{i=0}^{N}$
that act on gates executed at different times $\{t_{i}\}_{i=0}^{N}$.
If these gates are generated by $H(t)$, the inverse pulse $H_{I}(t)$
not only produces the corresponding inverse gates, but also reverses
their time order. Therefore, under $H_{I}(t)$ the noise mechanism
that acts at time $t_{j}$ is not $\mathcal{L}_{j}$ but $\mathcal{L}_{N-j}$.
Equation (\ref{eq:18 inverse dissipator}) captures this reverse ordering
in a time-continuous fashion. 

We note that Eqs. (\ref{eq:16 inverse driving}) and (\ref{eq:18 inverse dissipator})
describe the action of $H_{I}(t)$ and $\mathcal{L}_{I}(t)$ in a
time interval $(0,T)$. To construct the cycle $K_{I}K$, the driving
$H_{I}(t)$ must be implemented between $t=T$ and $t=2T$. Therefore,
when extending the master equation (\ref{eq:5 evol including noise and coh errors})
to cover the total interval $(0,2T)$, we must perform the replacements
$H_{L}(t)\rightarrow H_{I}(t-T)\otimes I-I\otimes H_{I}(t-T)^{\textrm{T}}$
and $\mathcal{L}(t)\rightarrow\mathcal{L}_{I}(t-T)$ for $T\leq t\leq2T$.
On the other hand, $\delta H_{L}(t)$ can feature an arbitrary time
dependence, without requiring any particular relation between the
coherent errors affecting the evolutions $K$ and $K_{I}$. 

\subsection{Evaluating the incoherent infidelity using multiple $K_{I}K$ cycles}

In this section, we present our procedure for assessing the incoherent
infidelity. We start by defining the quantities 
\begin{align}
R_{k} & :=\langle\langle\rho_{k}|\rho_{0}\rangle\rangle\nonumber \\
 & =\langle\langle \rho_{0}|{(K_{I}K)^{k}}|\rho_{0}\rangle\rangle,\label{eq:19 Rk}
\end{align}
where $|\rho_{k}\rangle\rangle=(K_{I}K)^{k}|\rho_{0}\rangle\rangle$, and 
\begin{align}
\sigma_{n} & :=\sum_{k=0}^{n}a_{k}^{(n)}\langle\langle \rho_{0}|{(K_{I}K)^{k}}|\rho_{0}\rangle\rangle\nonumber \\
 & =\sum_{k=0}^{n}a_{k}^{(n)}R_{k}\nonumber \\
 & \cong\sum_{k=0}^{n}a_{k}^{(n)}\langle\langle \rho_{0}|e^{k\chi}|\rho_{0}\rangle\rangle.\label{eq:20 sigma_n}
\end{align}
The third line follows from the approximation (\ref{eq:13 KIK approx-1}),
and we also define $\chi:=2\Omega_{1}-i\Theta$. In the case of a
pure initial state $\rho_{0}$, $R_{k}$ is the survival probability
measured after $k$ cycles $K_{I}K$. Namely, the probability that
the system is found in the initial state when the circuit $(K_{I}K)^{k}$
is executed. 

Assuming that the operator $\chi$ is diagonalizable, an eigenvalue
of $(K_{I}K)^{k}$ has the form $e^{kx}$, where $x$ is an eigenvalue
of $\chi$. Notice that $x$ is a complex number in general. Furthermore,
$\sum_{k=0}^{n}a_{k}^{(n)}(K_{I}K)^{k}$ has a corresponding eigenvalue
given by $\lambda(x):=\sum_{k=0}^{n}a_{k}^{(n)}e^{kx}$. For sufficiently
weak noise and coherent errors the cycle $K_{I}K$ is close to the
identity operator $I_{L}$, and therefore the norm of $x$ satisfies
$|x|\ll1$. Thus, we can approximate any eigenvalue $\lambda(x)$
by a truncated Taylor series around $x=0$. 

For reasons that will shortly be clarified, our goal is to approximate
$x$ through $\lambda(x)$. The quality of this approximation will
depend on the maximum number of cycles $n$ in the circuit combination
$\sum_{k=0}^{n}a_{k}^{(n)}(K_{I}K)^{k}$. To obtain this approximation,
we impose the conditions 
\begin{align}
\left(\frac{d^{j}}{dx^{j}}\lambda\right)_{x=0} & =0\:,\:j=0,2,3,...,n,\label{eq:21 deriv cond n-1}\\
\left(\frac{d}{dx}\lambda\right)_{x=0} & =1,\label{eq:22 deriv cond 1-1}
\end{align}
which lead to the truncated Taylor expansion 
\begin{eqnarray}
\lambda(x) & = & x+\frac{(-1)^{n+1}}{n+1}x{}^{n+1}+O\left(x^{n+2}\right).\label{eq:23 truncated Tay exp for lambda}
\end{eqnarray}

 By rescaling each eigenvalue $x$ as $x\rightarrow\varsigma x$, we can combine Eqs. (\ref{eq:20 sigma_n}) and (\ref{eq:23 truncated Tay exp for lambda}) to obtain 

\begin{align}
\sigma_{n} & \cong\left\langle \chi\right\rangle +\frac{(-1)^{n+1}}{n+1}\left\langle \chi^{n+1}\right\rangle +O\left(\varsigma^{n+2}\right)\nonumber \\
 & =2\left\langle \Omega_{1}\right\rangle +\frac{(-1)^{n+1}}{n+1}\left\langle \chi^{n+1}\right\rangle +O\left(\varsigma^{n+2}\right).\label{eq:24 sigma_n and purity loss}
\end{align}
In the second line of Eq. \eqref{eq:24 sigma_n and purity loss} we write $\left\langle \chi\right\rangle =2\left\langle \Omega_{1}\right\rangle $,
using the already mentioned property $\langle\langle \rho_{0}|\Theta|\rho_{0}\rangle\rangle=0$.
By combining Eq. (\ref{eq:24 sigma_n and purity loss}) with Eq. (\ref{eq:12 linear incoh infidelity}),
we arrive to our main result:
\begin{equation}
\varepsilon_{\textrm{inc}}(\varrho^{(\textrm{id})},\tilde{\varrho})\cong-\frac{\sigma_{n}}{2}+O\left(\varsigma^{n+1}\right)+O\left(\xi^{2}\right).\label{eq:24.1 incoh infidel and sigma_n}
\end{equation}

This implies that the incoherent infidelity can be estimated with higher
accuracy by increasing $n$. The measurement of
$\sigma_{n}$ involves implementing the circuits $(K_{I}K)^{k}$,
for $0\leq k\leq n$, measuring the corresponding quantities $R_{k}$,
and computing the weighted sum $\sum_{k=0}^{n}a_{k}^{(n)}R_{k}$. In practice, the convergence of $\sigma_{n}$ within a given experimental precision determines the maximum value of  $n$. This means that, once  $|\sigma_{n+1}-\sigma_{n}|$ is smaller than the target experimental uncertainty, it is not practical to keep increasing $n$.
For clarity, we also mention that the variance in the estimation of $\sigma_{n}$ can be reduced by performing a sufficient number of shots
(executions) of the circuits $(K_{I}K)^{k}$. 

\subsection{Explicit form of the coefficients $a_{k}^{(n)}$}

Let us now discuss the explicit form of the coefficients $a_{k}^{(n)}$
that are obtained from Eqs. (\ref{eq:21 deriv cond n-1}) and (\ref{eq:22 deriv cond 1-1}).
These equations are equivalent to the linear system  
\begin{align}
\sum_{k=0}^{n}k^{j}a_{k}^{(n)} & =0,\:j=0,2,3,...n,\label{eq:25 ak eq}\\
\sum_{k=0}^{n}ka_{k}^{(n)} & =1.\label{eq:26 ak eqk1}
\end{align}
Such a system can be conveniently written as 
\begin{equation}
(a_{0}^{(n)},a_{1}^{(n)},a_{2}^{(n)},..,a_{n}^{(n)})V_{n}=(0,1,0,..,0),\label{eq:27 eqs. in terms of Vandermonde matrix}
\end{equation}
 where 

\begin{equation}
V_{n}=\left(\begin{array}{ccccc}
1 & 0 & 0 & .. & 0\\
1 & 1^{1} & 1^{2} & .. & 1^{n}\\
1 & 2^{1} & 2^{2} & .. & 2^{n}\\
.. & .. & .. & .. & ..\\
1 & n & n^{2} & .. & n^{n}
\end{array}\right).\label{eq:28 Vandermonde matrix}
\end{equation}
is an $(n+1)\times(n+1)$ Vandermonde matrix. Accordingly, the solution
to the coefficients $a_{k}^{(n)}$ corresponds to the second row of
the inverse of $V_{n}$. If the elements $\left(V_{n}^{-1}\right)_{j,k}$
of this inverse are identified with indexes $1\leq j,k\leq n+1$,
we have that 

\begin{align}
a_{k}^{(n)} & =\left(V_{n}^{-1}\right){}_{2,k+1}.\label{eq:29 ak ivdm}
\end{align}

In particular, for $1\leq n\leq4$ we obtain: 
\begin{align}
\sigma_{1} & =-R_{0}+R_{1},\label{eq:30 sigma1}\\
\mathcal{\sigma}_{2} & =-\frac{3}{2}R_{0}+2R_{1}-\frac{1}{2}R_{2},\label{eq:31 sigma2}\\
\mathcal{\sigma}_{3} & =-\frac{11}{6}R_{0}+3R_{1}-\frac{3}{2}R_{2}+\frac{1}{3}R_{3},\label{eq:32 sigma3}\\
\mathcal{\sigma}_{4} & =-\frac{25}{12}R_{0}+4R_{1}-3R_{2}+\frac{4}{3}R_{3}-\frac{1}{4}R_{4}.\label{eq:33 sigma4}
\end{align}
As we shall see in Sec. \textcolor{blue}{V}, while $\mathcal{\sigma}_{1}=R_{1}-R_{0}$
turns out to be specially sensitive to coherent errors, $\sigma_{n\geq2}$
provides an accurate estimation of the incoherent infidelity.\textcolor{red}{{} }

\subsection{Robustness to SPAM errors }

SPAM errors are errors that occur in the preparation of the initial
state and in the measurement of the final state. A desirable characteristic
of any benchmarking tool is that it can separate the SPAM contribution
from the total error, in order to provide a reliable quantification
of the errors associated with the evolution, i.e. the circuit. We
show that the measurement of $\sigma_{n}$ is robust to SPAM errors
that are comparable to the errors encapsulated by $\chi$. In this
way, the estimation of the incoherent infidelity is also robust to
this kind of SPAM errors. 

Let $U_{\textrm{p}}$ be the ideal (error-free) evolution that is
used to prepare the pure state $|\rho_{0}\rangle\rangle$. If the fiducial
state is denoted by $|0\rangle\rangle$, we have that $|\rho_{0}\rangle\rangle=U_{\textrm{p}}|0\rangle\rangle$.
Note that while here we are not using the subindex $L$ to indicate
that $U_{\textrm{p}}$ is a unitary in Liouville space, this is understood
from the context. Similarly, an ideal measurement of the state $\rho_{0}$
is described by the row vector $\langle\langle \rho_{0}|=\langle\langle 0|U_{\textrm{p}}^{\dagger}$.
In the error-free scenario, $R_{k}=\langle\langle 0|U_{\textrm{p}}^{\dagger}(K_{I}K)^{k}U_{\textrm{p}}|0\rangle\rangle$
would be measured by executing the circuit $U_{\textrm{p}}^{\dagger}(K_{I}K)^{k}U_{\textrm{p}}$
and measuring the fiducial state $|0\rangle\rangle$. However, instead of
the perfect unitaries $U_{\textrm{p}}$ and $U_{\textrm{p}}^{\dagger}$
what is implemented in practice are error-prone circuits $K_{\textrm{p}}$
and $K_{\textrm{m}}$, where the subscripts $p$ and $m$ stand for
preparation and measurement, respectively. These evolutions can be
written as (see Appendix \textcolor{blue}{III})
\begin{align}
K_{\textrm{p}} & =U_{\textrm{p}}e^{\Omega_{\textrm{p}}},\label{eq:34 Kp}\\
K_{\textrm{m}} & =e^{\bar{\Omega}_{\textrm{m}}}U_{\textrm{p}}^{\dagger},\label{eq:35 Km}\\
\bar{\Omega}_{\textrm{m}} & =U_{\textrm{p}}^{\dagger}\Omega_{\textrm{m}}U_{\textrm{p}}
\end{align}
where $\Omega_{\textrm{p}}$ ($\Omega_{\textrm{m}}$) is the Magnus
expansion that characterizes preparation (measurement) errors. We
remark that Eqs. (\ref{eq:34 Kp}) and (\ref{eq:35 Km}) represent
the exact solutions for $K_{\textrm{p}}$ and $K_{\textrm{m}}$, since
they include the full Magnus expansions and not only the first Magnus
terms. Importantly, $\Omega_{\textrm{p}}$ and $\bar{\Omega}_{\textrm{m}}$
fully incorporate noise and coherent errors in the preparation and
measurement stages, and in general $\bar{\Omega}_{\textrm{m}}$ and
$\Omega_{\textrm{p}}$ are unrelated to each other. 

In this way, under SPAM errors the quantity that is measured when
trying to estimate $\sigma_{n}$ is 
\begin{align}
\sigma'_{n} & =\sum_{k=0}^{n}a_{k}^{(n)}R'_{k}\nonumber \\
 & =\sum_{k=0}^{n}a_{k}^{(n)}\langle\langle |K_{\textrm{m}}{\color{red}{\normalcolor \left(K_{I}K\right)^{k}}}K_{\textrm{p}}|0\rangle\rangle.\label{eq:36 Sigma_n'}
\end{align}
In Appendix \textcolor{blue}{III}, we show that 
\begin{equation}
|\sigma'_{n}-\sigma_{n}|=O\left(\left\Vert \Omega_{\textrm{m}}\right\Vert \left\Vert \chi\right\Vert +\left\Vert \Omega_{\textrm{p}}\right\Vert \left\Vert \chi\right\Vert \right),\label{eq:37 sigma_n' and sigma_n}
\end{equation}
where $\left\Vert \cdot\right\Vert $ denotes the spectral norm. This
follows from the fact that any contribution to $|\sigma'_{n}-\sigma_{n}|$
involves products of the form $\left\Vert \Omega_{\textrm{m}}\right\Vert ^{k}\left\Vert \Omega_{\textrm{p}}\right\Vert ^{l}\left\Vert \chi\right\Vert ^{m}$,
with $k+l\geq1$ and $m\geq1$. Since we assume weak errors $\left\Vert \chi\right\Vert ,\left\Vert \Omega_{\textrm{p}}\right\Vert ,\left\Vert \Omega_{\textrm{m}}\right\Vert \ll1$,
these contributions are negligible, and the dominant terms in the
difference between $\sigma'_{n}$ and the SPAM-free quantity $\sigma_{n}$
are the quadratic corrections $\left\Vert \Omega_{\textrm{m}}\right\Vert \left\Vert \chi\right\Vert $
and $\left\Vert \Omega_{\textrm{p}}\right\Vert \left\Vert \chi\right\Vert $.
Crucially, while each $R'_{k}$ has a correction term which is $O\left(\left\Vert \Omega_{\textrm{p}}\right\Vert ,\left\Vert \Omega_{\textrm{m}}\right\Vert \right)$,
these leading order corrections cancel out when calculating $\sigma_{n}'$
and the SPAM corrections enter only in second order. Finally, we remark
that Eq. (\ref{eq:37 sigma_n' and sigma_n}) holds whenever $\sigma'_{n}$
contains all the terms affected by SPAM errors, including $R'_{0}=\langle\langle 0|K_{\textrm{m}}K_{\textrm{p}}|0\rangle\rangle$.
Therefore, $R'_{0}$ must also be measured and cannot simply be replaced
by its ideal value $R_{0}=1$.\textcolor{blue}{{} }We have avoided  writing
1 instead of $R_{0}$ in Eqs. (\ref{eq:30 sigma1})-(\ref{eq:33 sigma4}),
to further emphasize this fact. 

In the case of readout errors, it is also possible to complement our
method with techniques that mitigate their impact and help to achieve
the regime of SPAM robustness characterized in Sec. \textcolor{blue}{IVE}. In particular,
Appendix \textcolor{blue}{IV} discusses how to mitigate
local readout errors in a simple and scalable manner. Such strategy
would allow us to efficiently estimate the incoherent infidelity in
large systems, where local readout errors may be substantial, so long
as correlated readout errors remain sufficiently small.

\section{Illustrative examples}

We illustrate now our technique for measuring the incoherent infidelity
using two numerical examples. In Fig. \textcolor{blue}{1}, we consider the incoherent
infidelity associated with noise in the circuit that generates a five-qubit
GHZ (Greenberger--Horne--Zeilinger) state. The second
example, illustrated in Fig. \textcolor{blue}{2}, refers to the evaluation of the \emph{average}
incoherent infidelity of a CNOT gate. While the example of Fig. \textcolor{blue}{1}
does not include SPAM errors, we do so in the second example. The
analysis of the GHZ state is mainly aimed to show the resilience of
our method to coherent errors, depicted in Fig. \textcolor{blue}{1} by the convergence
of our estimation to the actual value of $\varepsilon_{\textrm{inc}}(\varrho^{(\textrm{id})},\tilde{\varrho})$.
The second example involves the estimation of the gate incoherent
infidelity, obtained by averaging over a set of initial states. We
consider specifically input states generated by error-prone Clifford
gates, as well as SPAM errors in the fiducial state and readout errors.
This example demonstrates the robustness of our method to the associated
SPAM errors. In addition, it serves to illustrate how our technique
can also be used to evaluate the average incoherent error of a target
circuit. 

For simplicity, the unitary evolutions in this section will be written
using the traditional notation. That is, as unitary operators that
act on quantum states in Hilbert space. 

\subsection{Estimating the incoherent infidelity for the preparation of a five-qubit
GHZ state }

The five-qubit GHZ state can be prepared by using a circuit composed
of a Hadamard gate and four CNOT gates. We remark
that in this example the initial state is the fiducial state $|0\rangle$,
and the circuit that prepares the GHZ state represents the target
evolution that we want to assess. To simulate the execution of this
circuit in a real physical system, we consider the cross-resonance
interaction employed for the CNOT gate in some superconducting
qubits \cite{alexander2020qiskit}. Letting $\{X,Y,Z\}$ denote the
Pauli matrices, the cross resonance interaction between the qubits
$j$ and $k$ is given by $Z_{j}\otimes X_{k}$, where $Z_{j}$ is
the $Z$ Pauli matrix applied on qubit $j$, and $X_{k}$ is the $X$
Pauli matrix applied on qubit $k$. By combining it with local rotations,
this interaction leads to the following CNOT unitary
\begin{equation}
U_{CNOT}^{(j,k)}=e^{i\frac{\pi}{4}X_{k}}e^{-i\frac{\pi}{4}Z_{j}\otimes X_{k}}e^{i\frac{\pi}{4}Z_{j}},\label{eq:37.2 cross reson CNOT}
\end{equation}
where $j$ is the control qubit and $k$ is the target qubit.

The imperfect preparation of the GHZ state is simulated by including
errors that act during the cross resonance pulse. On the other hand,
we assume error-free implementations for the single-qubit rotations
$e^{i\frac{\pi}{4}X_{k}}$ and $e^{i\frac{\pi}{4}Z_{j}}$, and for
the Hadamard gate. Noise is characterized by dephasing and amplitude
damping channels that act identically on the five qubits used to prepare
the GHZ state, as described in Appendix \textcolor{blue}{V}. For the coherent errors, we consider the cross
talk term \cite{krinner2020benchmarking} 
\begin{equation}
\delta H^{(j,k)}=\eta Z_{j}\otimes Z_{k},\label{eq:37.10 ZZ coherent error}
\end{equation}
which affects the ideal cross resonance interaction $Z_{j}\otimes X_{k}$. The parameter $\eta$ characterizes the strength of this
error contribution. Note that since the interaction (\ref{eq:37.10 ZZ coherent error})
is not controllable, $\eta$ does not change its sign when implementing
the pulse inverse. 

We simulate the estimation of the incoherent infidelity by computing
$\sigma_{n}$ in Eq. (\ref{eq:20 sigma_n}). In this case, the circuit
$K$ corresponds to the error-prone circuit used to prepare the GHZ
state, and $K_{I}$ is the associated inverse. The incoherent infidelity
$\varepsilon_{\textrm{inc}}(\varrho^{(\textrm{id})},\tilde{\varrho})$
is estimated by $-\sigma_{n}/2$, according to Eq.
(\ref{eq:24.1 incoh infidel and sigma_n}). For  (incoherent) infidelities around 0.05, our simulations yield estimations of $\sim0.04$ using  $-\sigma_{n}/2$, for $n\sim5$. However, for the sake of demonstration, we consider smaller infidelities in the example of Fig. \textcolor{blue}{1}, such that $-\sigma_{n}/2$ virtually converges to $\varepsilon_{\textrm{inc}}$. 

The initial state $\rho_{0}$
is the five-qubit ground state (where all the qubits are in the ``0''
state), and, for simplicity of demonstration, we do not consider SPAM
errors. In this way, we can focus on the robustness of our method
to coherent errors, which is the main purpose of this simulation.
Moderate SPAM errors can be added without affecting the outcome of
the simulation. This is illustrated in the example presented in Sec.
\textcolor{blue}{VB}. 

\begin{figure}

\centering{}\includegraphics[scale=0.5]{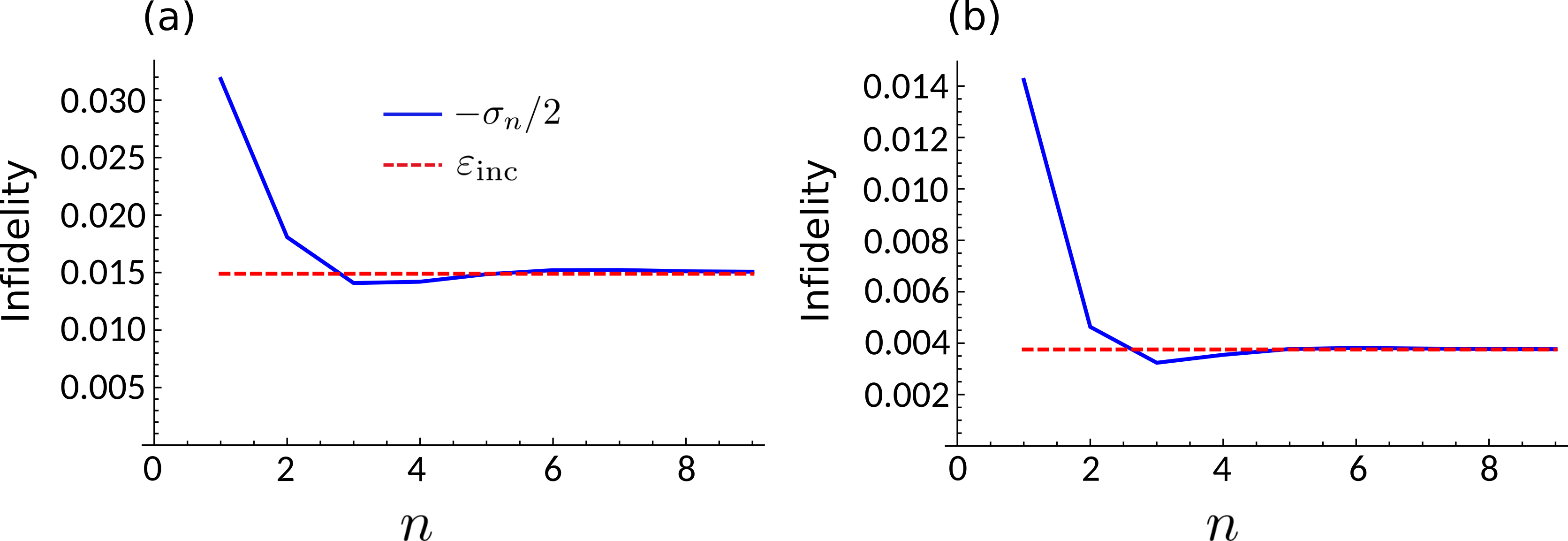}\caption{Incoherent infidelity for the implementation of a five-qubit GHZ state.
The dashed red lines depict the exact value and the solid blue lines
correspond to the estimation using $\sigma_{n}$ (cf. Eq. (\ref{eq:24.1 incoh infidel and sigma_n})),
as a function of the number of cycles $K_{I}K$ (denoted by $n$).
(a) Simulation parameters are $\eta T=0.0312$ and $\xi T=0.0035$,
where $T$ is the evolution time of the cross resonance interaction,
$\eta$ characterizes the strength of coherent errors, and $\xi$
characterizes the noise strength. (b) In this case, we set $\eta T=0.02$
and $\xi T=0.000351$. }
\end{figure}

The simulation parameters are chosen as follows. To obtain the ideal
CNOTs in the error-free case, we set the evolution time for the cross
resonance interaction to $T=\pi/4$. The coherent error parameter
$\eta$ is such that $\eta T=0.0312$, in Fig. \textcolor{blue}{1(a)}, and $\eta T=0.02$
for Fig. \textcolor{blue}{1(b)}. In Appendix \textcolor{blue}{V},
we also introduce a parameter $\xi$ that characterizes the strength
of the noise ($\xi=0$ corresponds to zero noise) and set it to $\xi T=0.0035$
and $\xi T=0.000351$ for Figs. \textcolor{blue}{1(a)} and \textcolor{blue}{1(b)}, respectively. These
values of $\eta$ and $\xi$ are low enough to maintain consistency
with the approximations intrinsic to our method. 

The results presented in Fig. \textcolor{blue}{1} show convergence of $-\sigma_{n}/2$
to the actual value of $\varepsilon_{\textrm{inc}}(\varrho^{(\textrm{id})},\tilde{\varrho})$
(dashed red lines), at approximately $n=5$ $K_{I}K$ cycles. Thus,
they provide numerical evidence of the accuracy achieved by our method,
in the regime of weak noise. In Sec. \textcolor{blue}{VI}, we precisely
define this noise regime and characterize the corresponding accuracy.
The big difference between $-\sigma_{1}/2$ and $\varepsilon_{\textrm{inc}}(\varrho^{(\textrm{id})},\tilde{\varrho})$
indicates that the linear order approximation (\ref{eq:15 linear approx to <KIK>})
is not sufficient to suppress the effect of coherent errors in the
calculation of $\varepsilon_{\textrm{inc}}(\varrho^{(\textrm{id})},\tilde{\varrho})$.
As $n$ increases, this effect is systematically mitigated, and we
can obtain an accurate estimation of the incoherent infidelity. We
also note that the relative error for the estimation using $\sigma_{1}$
is substantially larger in Fig. \textcolor{blue}{1(b)}. This can be explained by the
fact that the coherent error strength is not much smaller than the
corresponding to Fig. \textcolor{blue}{1(a)}, while the noise strength is reduced by
one order of magnitude. Therefore, the first-order ($n=1$) estimation
of the incoherent infidelity is more affected by coherent errors in
Fig. \textcolor{blue}{1(b)}. However, in both cases increasing $n$ fixes this problem
and leads to fast convergence to $\varepsilon_{\textrm{inc}}(\varrho^{(\textrm{id})},\tilde{\varrho})$.

\subsection{Estimating the average incoherent infidelity for a CNOT gate affected
by SPAM errors}

In this example, the target circuit is a CNOT gate which acts on qubits
labeled by $j=1$ and $k=2$. Rather than estimating the incoherent
infidelity associated with a single initial state $\rho_{0}$, we
consider the average over many randomly chosen (two-qubit) Clifford
gates employed for preparation and measurement. Specifically, in Fig.
\textcolor{blue}{2} we present results for $M=300$ different initial states obtained
from these gates. In methods like Interleaved RB, this averaging is
useful to tailor the noise into a global depolarizing channel, from
which it is possible to extract the error per gate. However, such
a noise tailoring can be hampered by coherent errors in the Clifford
gates used for preparation and measurement. While it has been shown
that this can cause an incorrect estimation of the error rate \cite{proctor2017randomized}, in the following
example we illustrate that this is not the case with the present method. 

The average incoherent infidelity is defined as 

\begin{equation}
\bar{\varepsilon}_{\textrm{inc}}:=\frac{1}{M}\sum_{m=1}^{M}\varepsilon_{\textrm{inc}}(\varrho_{m}^{(\textrm{id})},\tilde{\varrho}_{m}),\label{eq:37.25 average inc infid}
\end{equation}
where $\varrho_{m}^{(\textrm{id})}$ and $\tilde{\varrho}_{m}$ are
states corresponding to the $m$th preparation. More precisely, $\varrho_{m}^{(\textrm{id})}$
is the error-free final state and $\tilde{\varrho}_{m}$ is the state
obtained when the target evolution only contains noise and no coherent
errors. We also stress that, for capturing the errors associated with
the gate, SPAM errors are absent in the computation of the exact infidelity
(\ref{eq:37.25 average inc infid}), as per the definition given in
Eq. (\ref{eq:6 incoh infid}). However, they are included in the simulated
estimation of the average incoherent infidelity. 

To simulate the estimation of $\bar{\varepsilon}_{\textrm{inc}}$,
we include coherent errors and noise that afflict $U_{CNOT}^{(1,2)}$
as well as the Clifford gates used for preparation and measurement.
SPAM errors contain also a contribution that is relevant for quantum
computing architectures such as the one used by IBM. Concretely, we
refer to readout errors \cite{maciejewski2020mitigation,nation2021scalable} and potential errors in
the preparation of the fiducial state $|0\rangle$ \cite{landa2022experimental}. Our modeling of these errors
is based on Reference \cite{landa2022experimental},
and is described in Appendix \textcolor{blue}{V}. In combination with
the errors affecting the Clifford gates for preparation and measurement,
they constitute the SPAM errors for the simulation presented below. 

All the CNOT gates are subjected to the coherent error (\ref{eq:37.10 ZZ coherent error}),
and we shall use the angle $\phi=\eta T$ to characterize the associated
error strength. In addition, following Reference \cite{proctor2017randomized}, we introduce another coherent error that affects single-qubit
gates. Specifically, let $R_{W}(\alpha)=e^{-i\frac{\alpha}{2}W}$
denote a single-qubit rotation by an angle $\alpha$, around some
``axis'' $W\in\{X,Y,Z\}$ of the Bloch sphere. We consider the following
modification of the error-free rotations $R_{W}(\pm\pi/2)$, for $W=X,Y$
\cite{proctor2017randomized}: 

{
\begin{align}
R_{W}(\pi/2) & \rightarrow e^{-i\theta Z}R_{W}(\pi/2),\label{eq:37.26 single-qub coh error 1}\\
R_{W}(-\pi/2) & \rightarrow R_{W}(-\pi/2)e^{i\theta Z}.\label{eq:37.27 single-qub coh error 2}
\end{align}
}Importantly, any single-qubit Clifford gate can be compiled using
sequences of the ideal gates $R_{X}(\pm\pi/2)$ and $R_{Y}(\pm\pi/2)$.
Furthermore, it was shown in \cite{proctor2017randomized} that
RB is sensitive to the coherent error described by Eqs. (\ref{eq:37.26 single-qub coh error 1})
and (\ref{eq:37.27 single-qub coh error 2}). This error is added
to any single-qubit Clifford gate in our simulation, including the
$R_{X}$ gate in the target CNOT. Regarding noise, only the cross
resonance pulse (cf. Eq. (\ref{eq:37.2 cross reson CNOT})) in the
CNOT gates is affected by local dephasing and amplitude damping. A
complete description of all these error sources can be found in Appendix \textcolor{blue}{V}.

In Fig. \textcolor{blue}{2}, we plot $-\sigma_{1}^{\prime}/2$ (solid red curves) and
$-\sigma_{2}^{\prime}/2$ (solid purple curves) as a function of the
angle $\phi$ , for values $\phi\in\{0,0.01,0.02,0.03,0.04,0.05\}$.
Moreover, the noise strength is chosen such that $\xi T=0.001$. In
addition to the incoherent infidelity, we also apply Interleaved RB \cite{magesan2012efficient}
to estimate the average infidelity of the target CNOT gate. Details
about this estimation are given in Appendix \textcolor{blue}{VI}. As
opposed to our method, the infidelity obtained via Interleaved RB
aims to encompass all the error sources and not only the contribution
from noise. This means that, instead of $\varepsilon_{\textrm{inc}}(\varrho_{m}^{(\textrm{id})},\tilde{\varrho}_{m})$,
the infidelity corresponding to the $m$-th preparation is \textcolor{red}{${\normalcolor \varepsilon(\varrho_{m}^{(\textrm{id})},\varrho_{m})=1-F(\varrho_{m}^{(\textrm{id})},\varrho_{m})}$},
where $\varrho_{m}$ is the actual final state that results from the
application of $K$. Thus, the estimation of 
\begin{equation}
\bar{\varepsilon}:=\frac{1}{M}\sum_{m=1}^{M}\varepsilon(\varrho_{m}^{(\textrm{id})},\varrho_{m})\label{eq:37.32 average total infidelity}
\end{equation}
is the relevant reference for evaluating the performance of Interleaved
RB. 

\begin{figure}

\centering{}\includegraphics[scale=0.6]{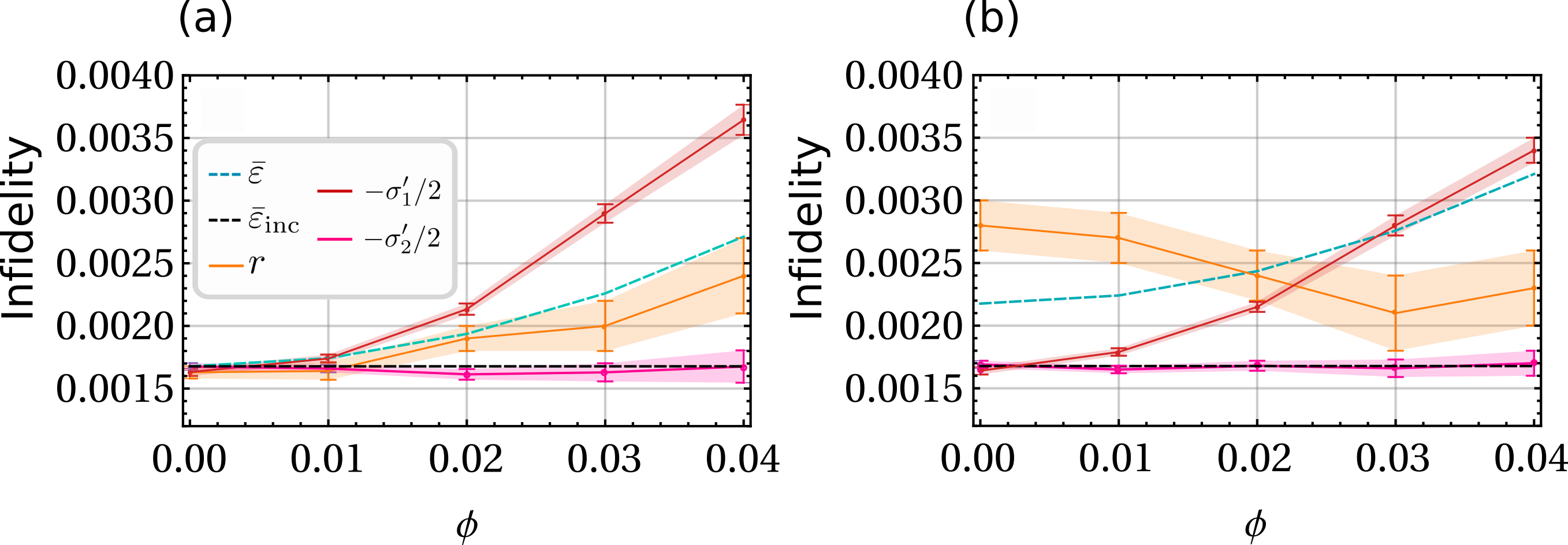}\caption{Average incoherent infidelity and average infidelity for the implementation
of a CNOT gate. These averages are taken over $M=300$
two-qubit Clifford gates, employed for the preparation of the initial
state. In these plots, $\bar{\varepsilon}$ is the exact infidelity
(\ref{eq:37.32 average total infidelity}), $r$ is the estimate of
$\bar{\varepsilon}$ using Interleaved RB, and $\bar{\varepsilon}_{\textrm{inc}}$
is the exact incoherent infidelity (\ref{eq:37.25 average inc infid}).
Moreover, $-\sigma_{1}^{\prime}/2$ and $-\sigma_{2}^{\prime}/2$
are estimates of $\bar{\varepsilon}_{\textrm{inc}}$ that involve
SPAM errors. The error bars correspond to one standard deviation when
only one measurement is performed per each Clifford gate. In (a) and
(b), the single-qubit coherent error is set to $\theta=0$ and $\theta=0.05$,
respectively. }
\end{figure}

The angle $\theta$ is set to $\theta=0$ and $\theta=0.05$ in Figs.
\textcolor{blue}{2(a)} and \textcolor{blue}{2(b)}, respectively. In both cases, we observe that the estimation
of $\bar{\varepsilon}$ via Interleaved RB (solid orange curves) significantly
deviates from the exact value (dashed cyan curves). This is specially
noticeable in Fig. \textcolor{blue}{2(b)}, where the deviation is clear even for $\phi=0$.
Such a mismatch is an indication that either coherent errors are not
properly addressed in the  Interleaved RB method, or that the error
rate that it measures cannot always be associated with the infidelity. On the other hand, Fig. \textcolor{blue}{2} shows that $-\sigma_{2}^{\prime}/2$
(solid magenta curves) approaches well the incoherent infidelity $\bar{\varepsilon}_{\textrm{inc}}$(dashed
black curves), irrespective of the type of coherent errors included.
We also emphasize that SPAM errors are fully taken into account in
the simulation of $-\sigma_{1}^{\prime}/2$ (solid red curves) and
$-\sigma_{2}^{\prime}/2$.\textcolor{purple}{{} }Thus, the proximity
of $-\sigma_{2}^{\prime}/2$ and $\bar{\varepsilon}_{\textrm{inc}}$
illustrates the resilience of our estimation to SPAM errors. Finally,
we remark that, similarly to Fig. \textcolor{blue}{1}, the linear approximation $-\sigma_{1}^{\prime}/2$
fails to provide a good estimate of the incoherent infidelity. However,
with only two $K_{I}K$ cycles we obtain the accurate estimation observed
in Fig. \textcolor{blue}{2}, which is specially evident by the overlap between the magenta
curve and the dashed black curve in Fig. \textcolor{blue}{2(b)}. 

To conclude this section, we point out that for each initial state
the measurement of $\sigma_{n}^{\prime}$ yields an estimate of the
incoherent infidelity associated with that specific input. This contrasts
with Interleaved RB and other benchmarking tools for target circuits,
where measurement data can only provide estimates of the circuit infidelity. In other words, although Eq. (\ref{eq:37.32 average total infidelity})
is defined in terms of the single-realization quantities $\varepsilon(\varrho_{m}^{(\textrm{id})},\varrho_{m})$,
it is not possible to assess these infidelities via Interleaved RB.
Only the average $\bar{\varepsilon}$ can be estimated. Our method
is not restricted by this limitation and could also be useful to learn
about the noise fluctuations associated with different initial conditions.
In particular, by comparing different infidelities $\varepsilon_{\textrm{inc}}(\varrho_{m}^{(\textrm{id})},\tilde{\varrho}_{m})$
between each other and with the average infidelity $\bar{\varepsilon}_{\textrm{inc}}$,
one could study how input sates influence circuit performance and
use this information for the design of effective error suppression
strategies. 

\section{Conditions for the scalability of the incoherent infidelity estimation}

In this section, we provide error bounds for the approximations (\ref{eq:12 linear incoh infidelity})
and (\ref{eq:13 KIK approx-1}). We argue that if the relevant error
parameters are small (as required for the validity of our approximations),
our method is scalable within the accuracy established by these bounds.
Specifically, we require that the accumulated noise and coherent errors
should remain sufficiently low as the circuit size increases. This
is a valid prerequisite for performing useful computations in the
NISQ era, where a systematic application of error correction is not yet available \cite{endo2021hybrid}. To characterize this condition in our formalism,
we set limits to the quantities $\intop_{0}^{T}dt\left\Vert \mathcal{L}_{\textrm{ext}}(t)\right\Vert $
and $\intop_{0}^{2T}dt\left\Vert \mathcal{L}_{\textrm{ext}}(t)-i\delta H_{L}(t)\right\Vert $,
introduced in Eqs. (\ref{eq:41 error bound 1}) and (\ref{eq:42 error bound 2}). 

For clarity, we are going to denote by $\Omega(T)$ and $\Omega(2T)$
the Magnus expansions associated with $K$ and $K_{I}K$, respectively.
In addition, we write the dissipator that characterizes the noise
in the extended time interval $(0,2T)$ as $\mathcal{L}_{\textrm{ext}}(t)$.
Therefore, the equation that governs the cycle $K_{I}K$ reads 
\begin{equation}
d_{t}|\rho\rangle\rangle=\left(-iH_{L,\textrm{ext}}(t)-i\delta H_{L}(t)+\mathcal{L}_{\textrm{ext}}(t)\right)|\rho\rangle\rangle,\label{eq:38 evolution eq for (0,2T)}
\end{equation}
where 
\begin{equation}
H_{L,\textrm{ext}}(t)=\begin{cases}
\begin{array}{c}
H_{L}(t),\textrm{ for \ensuremath{t\in(0,T)}},\\
H_{I}(t-T)\otimes I-I\otimes H_{I}^{\textrm{T}}(t-T),\textrm{ for \ensuremath{t\in(T,2T)}},
\end{array}\end{cases}\label{eq:39 extended driving}
\end{equation}
and 

\begin{equation}
\mc L_{\textrm{ext}}(t)=\begin{cases}
\begin{array}{c}
\mc L(t),\textrm{ for \ensuremath{t\in(0,T)}},\\
\mc L_{I}(t-T),\textrm{ for \ensuremath{t\in(T,2T)}.}
\end{array}\end{cases}\label{eq:40 extend dissipator}
\end{equation}

The error bounds presented in Eqs. (\ref{eq:41 error bound 1}) and
(\ref{eq:42 error bound 2}) are derived in Appendix \textcolor{blue}{
VII}. In the case of Eq. (\ref{eq:42 error bound 2}), we have a bound
that quantifies how close the approximation $e^{\chi}$ is to the
exact evolution $K_{I}K=e^{\Omega(2T)}$. The accuracy of this approximation
is directly related to the accuracy for estimating the incoherent
infidelity using $K_{I}K$ cycles. More precisely, note that the right
hand side of Eq. (\ref{eq:24 sigma_n and purity loss}) is based on
Eq. (\ref{eq:13 KIK approx-1}), but the definition of $\sigma_{n}$
(first line of Eq. (\ref{eq:20 sigma_n})) and its estimation rely
on circuits that contain the actual cycle $K_{I}K$. Therefore, the
closer $K_{I}K$ to $e^{\chi}$, the better the estimation of $\varepsilon_{\textrm{inc}}(\varrho^{(\textrm{id})},\tilde{\varrho})$
by using $\sigma_{n}$. We also emphasize that since many inequalities
are involved in the derivation of (\ref{eq:41 error bound 1}) and
(\ref{eq:42 error bound 2}), it is probable that these bounds substantially
overestimate the actual errors. As a result, the regime of validity
of our method is expected to be larger than predicted by these bounds. 

The error in approximating the incoherent infidelity by (\ref{eq:12 linear incoh infidelity})
is given by $\left|\varepsilon_{\textrm{inc}}(\varrho^{(\textrm{id})},\tilde{\varrho})+\left\langle \Omega_{1}\right\rangle \right|$.
Importantly, this error quantifies the difference between $-\left\langle \Omega_{1}\right\rangle $
and the \textit{exact} incoherent infidelity $\varepsilon_{\textrm{inc}}(\varrho^{(\textrm{id})},\tilde{\varrho})=1-\langle\langle \rho_{0}|e^{\Omega(T)}|\rho_{0}\rangle\rangle$,
expressed in terms of the \textit{full} Magnus expansion $\Omega(T)$.
In Appendix \textcolor{blue}{VII}, we derive the bound 
\begin{equation}
\left|\varepsilon_{\textrm{inc}}(\varrho^{(\textrm{id})},\tilde{\varrho})+\left\langle \Omega_{1}\right\rangle \right|\leq e^{\intop_{0}^{T}dt\left\Vert \mathcal{L}_{\textrm{ext}}(t)\right\Vert }-\intop_{0}^{T}dt\left\Vert \mathcal{L}_{\textrm{ext}}(t)\right\Vert -1.\label{eq:41 error bound 1}
\end{equation}
The quantity $\intop_{0}^{T}dt\left\Vert \mathcal{L}_{\textrm{ext}}(t)\right\Vert =\intop_{0}^{T}dt\left\Vert \mathcal{L}(t)\right\Vert $
in this bound can be interpreted as the total noise affecting the
evolution $K$, and is such that $\intop_{0}^{T}dt\left\Vert \mathcal{L}(t)\right\Vert=O(\xi) \ll1$
in the weak noise regime. In particular, for $\intop_{0}^{T}dt\left\Vert \mathcal{L}(t)\right\Vert \leq0.05$
we have that $\left|\varepsilon_{\textrm{inc}}(\varrho^{(\textrm{id})},\tilde{\varrho})+\left\langle \Omega_{1}\right\rangle \right|\leq0.001$,
according to Eq. (\ref{eq:41 error bound 1}). 

Crucially, an expansion of the right hand side of
Eq. (\ref{eq:41 error bound 1}) shows that this bound is $O\left(\xi^{2}\right)$.
On the other hand, in the regime of weak noise the incoherent infidelity
is linear in $\mathcal{L}(t)$, as per Eqs. (\ref{eq:9 Omega1})-(\ref{eq:12 linear incoh infidelity}).
This indicates that, in this regime,
the error in the estimation of $\varepsilon_{\textrm{inc}}(\varrho^{(\textrm{id})},\tilde{\varrho})$
is substantially smaller than the estimated incoherent infidelity.

In the case of Eq. (\ref{eq:42 error bound 2}), we quantify the error
using the spectral norm of the difference between $K_{I}K$ and the
approximation $e^{2\Omega_{1}-i\Theta}$ in Eq. (\ref{eq:13 KIK approx-1}).
Since the error-free evolution corresponding to $K_{I}K$ is the identity
operator, we have that $K_{I}K=e^{\Omega(2T)}$. Is is also worth
noting that $\chi=2\Omega_{1}-i\Theta$ is the first Magnus term of
the expansion $\Omega(2T)$, as explained in Appendix
\textcolor{blue}{I}. The relevant error bound is given by 
\begin{equation}
\left\Vert e^{\Omega(2T)}-e^{\chi}\right\Vert \leq2\left(e^{\intop_{0}^{2T}dt\left\Vert \mathcal{L}_{\textrm{ext}}(t)-i\delta H_{L}(t)\right\Vert }-\intop_{0}^{2T}dt\left\Vert \mathcal{L}_{\textrm{ext}}(t)-i\delta H_{L}(t)\right\Vert -1\right).\label{eq:42 error bound 2}
\end{equation}

Here, we identify $\intop_{0}^{2T}dt\left\Vert \mathcal{L}_{\textrm{ext}}(t)-i\delta H_{L}(t)\right\Vert $
with the total evolution error for $K_{I}K$, including the coherent
contribution from $\delta H_{L}(t)$. Keeping in mind that $K_{I}K$
takes twice the time invested by $K$, we assume that $\intop_{0}^{2T}dt\left\Vert \mathcal{L}_{\textrm{ext}}(t)-i\delta H_{L}(t)\right\Vert \gtrsim2\intop_{0}^{T}dt\left\Vert \mathcal{L}_{\textrm{ext}}(t)\right\Vert $.
Thus, as maximum reference value for $\intop_{0}^{2T}dt\left\Vert \mathcal{L}_{\textrm{ext}}(t)-i\delta H_{L}(t)\right\Vert $
we consider $\intop_{0}^{2T}dt\left\Vert \mathcal{L}_{\textrm{ext}}(t)-i\delta H_{L}(t)\right\Vert =0.1$,
keeping in mind that before we assumed $\intop_{0}^{T}dt\left\Vert \mathcal{L}_{\textrm{ext}}(t)\right\Vert \leq0.05$.
By inserting this error limit into Eq. (\ref{eq:42 error bound 2}),
we find that $\left\Vert e^{\Omega(2T)}-e^{\chi}\right\Vert \leq0.01$.
In analogy with Eq. (\ref{eq:41 error bound 1}), we also remark that
the dominant contribution to the bound (\ref{eq:42 error bound 2})
is the quadratic term $2\left(\intop_{0}^{2T}dt\left\Vert \mathcal{L}_{\textrm{ext}}(t)-i\delta H_{L}(t)\right\Vert \right)^{2}$. 

\section{Conclusions }

In this paper, we have developed a method for estimating
the impact of noise on the fidelity of a target quantum evolution.
To this end, we introduced the ``incoherent infidelity'', as a measure
of the infidelity that excludes the contribution from coherent errors
affecting the tested circuit. Our technique is designed to evaluate
the performance of a given target evolution, with respect to a given
initial state, or as an average over a set of initial states. These
features are specially relevant if the goal is to benchmark a certain
quantum algorithm or quantum computation, rather than the overall
performance of a quantum device. In particular, our estimation of
the incoherent infidelity can be applied to quantum algorithms whose
success depends on validating the quality of some specific output,
or outputs.\textcolor{blue}{{} }

We presented numerical simulations that highlight different aspects
of our estimation of the incoherent infidelity. By simulating errors
in the preparation of a five-qubit GHZ state, we illustrated its robustness
to coherent errors. Another example involved the estimation of the
incoherent infidelity for a CNOT gate. In this case,
we computed the average infidelity over a set of random initial states
prepared by Clifford gates. We have shown analytically that the incoherent
infidelity estimation is also resilient to SPAM errors of moderate
magnitude, and corroborated this property in our simulation. We modeled
SPAM errors by considering an imperfect fiducial state and readout
errors, as well as noise and coherent errors in the Clifford gates
employed for preparation and measurement. 

Our method does not require the classical simulation of the error-free
state generated by the target circuit. Furthermore, there is no restriction
on the target evolution and the gates that compose it. It also admits
a simple experimental implementation, and the number of measurements
to achieve a certain accuracy is independent of the size of the system.
Therefore, it provides a scalable tool to assess the incoherent infidelity
in the weak noise regime, where our theory is valid. 

An open problem is the extension of this technique to cover larger
error rates. However, it is worth keeping in mind that, in the absence
of quantum error correction, the errors accumulated in quantum computers
must be low enough to allow useful computations. Under this restriction,
we believe that our technique can be used to benchmark quantum algorithms
with the potential of a practical quantum advantage. We also anticipate
that it can be combined with other benchmarking tools for obtaining
additional information about different classes of errors. For example,
it might be possible to estimate the strength of coherent errors if
the incoherent infidelity is complemented with estimates of the total
infidelity (as proposed e.g. in Reference \cite{proctor2022establishing}). 

\section{Acknowledgments}
Raam Uzdin is grateful for support
from the Israel Science Foundation (Grant No. 2556/20).

\section{Data availability statement}
All the information relevant for our simulations is included in the plots within the main text and the appendices.  

\bibliographystyle{apsrev}
\bibliography{Refs.bib}

\section*{Appendix I - Approximate evolutions using the Magnus expansion }

In this appendix, we derive the approximations given in Eqs. (\ref{eq:8  K without coh errors})
and (\ref{eq:13 KIK approx-1}). Let us start with the operator $\tilde{K}$,
which represents the solution to the equation 

\begin{equation}
d_{t}|\rho\rangle\rangle=\left(-iH_{L}(t)+\mc L(t)\right)|\rho\rangle\rangle.\label{eq:S1.1}
\end{equation}
Using the solution $U_{L}(t)$ to the error-free equation $d_{t}|\rho\rangle\rangle=-iH_{L}(t)|\rho\rangle\rangle$,
we define the interaction picture vector 
\begin{equation}
|\rho_{int}\rangle\rangle=U_{L}^{\dagger}(t)|\rho\rangle\rangle.\label{eq:S1.2}
\end{equation}
In this way, Eq. (\ref{eq:S1.1}) can be rewritten in the interaction
picture as 

\begin{equation}
d_{t}|\rho_{int}\rangle\rangle=\mc L^{int}(t)|\rho_{int}\rangle\rangle,\label{eq:S1.3}
\end{equation}
where $\mc L^{int}(t)=U_{L}^{\dagger}(t)\mc L(t)U_{L}(t)$. The general
solution to this equation can be expressed using the Magnus expansion
\cite{blanes2009magnus}. Using this approach, the state $|\rho_{int}(T)\rangle\rangle$
at time $T$ reads 
\begin{equation}
|\rho_{int}(T)\rangle\rangle=e^{\Omega(T)}|\rho_{int}(0)\rangle\rangle,\label{eq:S1.4}
\end{equation}
where $\Omega(T)=\sum_{i=1}^{\infty}\Omega_{i}(T)$ is the Magnus
expansion. The Magnus expansion is applicable to the general differential
equation $d_{t}|\rho_{int}\rangle\rangle=\Lambda(t)|\rho_{int}\rangle\rangle,$
with the operators $\Omega_{i}(T)$ expressed in terms of the operator
$\Lambda(t)$. For $\Lambda(t)=\mc L^{int}(t)$, the first two terms
$\Omega_{1}(T)$ and $\Omega_{2}(T)$ are given by 
\begin{align}
\Omega_{1}(T) & =\int_{0}^{T}dt\mc L^{int}(t),\label{eq:S1.5}\\
\Omega_{2}(T) & =\frac{1}{2}\int_{0}^{T}dt'\int_{0}^{t'}[\mc L^{int}(t'),\mc L^{int}(t)]dt.\label{eq:S1.6}
\end{align}

By multiplying both sides of Eq. (\ref{eq:S1.4}) by $U_{L}(T)$, we
obtain
\begin{equation}
|\rho(T)\rangle\rangle=U_{L}(T)e^{\Omega(T)}|\rho_{0}\rangle\rangle,\label{eq:S1.7}
\end{equation}
where $|\rho_{0}\rangle\rangle=|\rho_{int}(0)\rangle\rangle$ is the initial state
in the Schrodinger picture. Thus, Eq. (\ref{eq:8  K without coh errors})
is the result of approximating the exact solution $\tilde{K}:=U_{L}(t)e^{\Omega(T)}$,
by discarding the Magnus terms $\Omega_{i\geq2}(T)$. We also point
out that in the main text we write $\Omega_{1}(T)$ as $\Omega_{1}$. 

For the operator $K_{I}K$, we must find the solution to the equation
\begin{equation}
d_{t}|\rho\rangle\rangle=\left(-iH_{L,\textrm{ext}}(t)+\mathcal{L}_{\textrm{ext}}^{\prime}(t)\right)|\rho\rangle\rangle,\label{eq:S1.8}
\end{equation}
where 
\begin{equation}
H_{L,\textrm{ext}}(t)=\begin{cases}
\begin{array}{c}
H_{L}(t),\textrm{ for \ensuremath{t\in(0,T)}},\\
H_{I}(t-T)\otimes I-I\otimes H_{I}^{\textrm{T}}(t-T),\textrm{ for \ensuremath{t\in(T,2T)}},
\end{array}\end{cases}\label{eq:S1.9}
\end{equation}
and 

\begin{equation}
\mathcal{L}_{\textrm{ext}}^{\prime}(t)=\begin{cases}
\begin{array}{c}
\mc L^{\prime}(t),\textrm{ for \ensuremath{t\in(0,T)}},\\
-i\delta H_{L}(t)+\mc L_{I}(t-T),\textrm{ for \ensuremath{t\in(T,2T)}.}
\end{array}\end{cases}\label{eq:S1.10}
\end{equation}
We note that Eq. (\ref{eq:S1.8}) is an extension of Eq. (\ref{eq:5 evol including noise and coh errors})
to the total time interval $(T,2T)$. For $t\in(T,2T)$, $H_{L,\textrm{ext}}(t)$
and $\mathcal{L}_{\textrm{ext}}^{\prime}(t)$ are defined according
to the prescriptions given in Sec. \textcolor{blue}{IVB} of the main
text. Here, $\mc L^{\prime}(t)=-i\delta H_{L}(t)+\mc L(t)$ (cf. Eq.
(\ref{eq:5 evol including noise and coh errors})). We further remark
that since we consider general coherent errors in the time domains
$(0,T)$ and $(T,2T)$, the notation $\delta H_{L}(t)$ can be extended
to $(T,2T)$ without risk of confusion. 

Since Eq. (\ref{eq:S1.8}) is also of the general form $d_{t}|\rho_{int}\rangle\rangle=\Lambda(t)|\rho_{int}\rangle\rangle$, we
can resort to the Magnus expansion to express its solution. The interaction
picture vector $|\rho_{int}\rangle\rangle$ for $t\in(0,T)$ is defined through
Eq. (\ref{eq:S1.2}). For $t\in(T,2T)$, we must consider the ideal
evolution associated with this interval. To obtain the corresponding unitary, let us first consider the evolution $U_{I}(t)$ generated by the driving $H_{I}(t)$. For $t\in(0,T)$, $H_{I}(t)$ is such that it undoes the portion of the $U(T)$ generated in the time subinterval $(T-t,T)$. That is, $U_{I}(t)U(T)=U(T-t)$. Expressing this relationship in the interval $(T,2T)$ involves the time shift $t\rightarrow t-T$, which leads to $U(T-t)\rightarrow U(2T-t)$. Similarly, in Liouville space $U_{L}(T-t)\rightarrow U_{L}(2T-t)$. Therefore,

\begin{equation}
|\rho_{int}\rangle\rangle=\begin{cases}
\begin{array}{c}
U_{L}^{\dagger}(t)|\rho\rangle\rangle,\textrm{ for \ensuremath{t\in(0,T)}},\\
U_{L}^{\dagger}(2T-t)|\rho\rangle\rangle,\textrm{ for \ensuremath{t\in(T,2T)}.}
\end{array}\end{cases}\label{eq:S1.11}
\end{equation}

We stress that none of the derivations in this appendix or the rest of the present article assume some special structure of the Hamiltonian. The time dependence of $H(t)$ is general and commutativity at different times is not required. However, it is also important to keep in mind that $H_{I}(t)$ is by construction related to $H(t)$. Thus, the evolution can be arbitrary in the time interval $t\in(0,T)$, but not in the total interval $t\in(0,2T)$.  

Using Eq. (\ref{eq:S1.11}), the dynamical equation (\ref{eq:S1.8})
in the interaction picture reads 

\begin{equation}
d_{t}|\rho_{int}\rangle\rangle=\left(\mathcal{L}_{\textrm{ext}}^{\prime}\right)^{int}(t)|\rho_{int}\rangle\rangle,\label{eq:S1.11.1}
\end{equation}
where 

\begin{equation}
\left(\mathcal{L}_{\textrm{ext}}^{\prime}\right)^{int}(t)=\begin{cases}
\begin{array}{c}
U_{L}^{\dagger}(t)\mathcal{L}_{\textrm{ext}}^{\prime}U_{L}(t),\textrm{ for \ensuremath{t\in(0,T)}},\\
U_{L}^{\dagger}(2T-t)\mathcal{L}_{\textrm{ext}}^{\prime}U_{L}(2T-t),\textrm{ for \ensuremath{t\in(T,2T)}.}
\end{array}\end{cases}\label{eq:S1.12}
\end{equation}
The first Magnus term for $K_{I}K$ is obtained by integrating $\left(\mathcal{L}_{\textrm{ext}}^{\prime}\right)^{int}$
from $t=0$ to $t=2T$, and is given by 
\begin{align}
\Omega_{1}(2T) & =\int_{0}^{2T}dt\left(\mathcal{L}_{\textrm{ext}}^{\prime}\right)^{int}(t)\nonumber \\
 & =\int_{0}^{T}dtU_{L}^{\dagger}(t)\mathcal{L}_{\textrm{ext}}^{\prime}(t)U_{L}(t)+\int_{T}^{2T}dtU_{L}^{\dagger}(2T-t)\mathcal{L}_{\textrm{ext}}^{\prime}(t)U_{L}(2T-t).\label{eq:S1.13}
\end{align}

In contrast with Eq. (\ref{eq:S1.4}), $\left(\mathcal{L}_{\textrm{ext}}^{\prime}\right)^{int}(t)$
contains not only the contribution from noise, embodied by the dissipators
$\mc L(t)$ and $\mc L_{I}(t-T)$, but also the coherent errors $\delta H_{L}(t)$.
The total coherent error in $\Omega_{1}(2T)$ is $-i\Theta$, with
$\Theta$ expressed by Eq. (\ref{eq:14 accumulated coh error-1}).
Regarding the noise, we have the sum of the two terms $\int_{0}^{T}dtU_{L}^{\dagger}(t)\mathcal{L}(t)U_{L}(t)$
and $\int_{T}^{2T}dtU_{L}^{\dagger}(2T-t)\mc L_{I}(t-T)U_{L}(2T-t)$.
Since $\mc L_{I}(t-T)=\mc L(2T-t)$ (cf. Eq. (\ref{eq:18 inverse dissipator})),
we can perform the change of variable $t'=2T-t$, to obtain 
\begin{equation}
\int_{T}^{2T}dtU_{L}^{\dagger}(2T-t)\mc L_{I}(t-T)U_{L}(2T-t)=\int_{0}^{T}dt'U_{L}^{\dagger}(t')\mc L(t')U_{L}(t').\label{eq:S1.14}
\end{equation}
Therefore, the noisy component of $\Omega_{1}(2T)$ is given by $2\int_{0}^{T}dtU_{L}^{\dagger}(t)\mathcal{L}(t)U_{L}(t)=2\Omega_{1}(T)$.
This leads us to conclude that $K_{I}K\approx e^{2\Omega_{1}-i\Theta}$.

\section*{Appendix II - Proof of the property $\langle\langle\rho_{0}|\Theta|\rho_{0}\rangle\rangle=0$}

From the definition of $\Theta$ (cf. Eq. (\ref{eq:14 accumulated coh error-1})),
we have that 

\begin{equation}
\langle\langle\rho_{0}|\Theta|\rho_{0}\rangle\rangle=\intop_{0}^{T}\langle\langle\rho(t)|\delta H_{L}(t)|\rho(t)\rangle\rangle dt+\intop_{T}^{2T}\langle\langle\rho(2T-t)|\delta H_{L}(t)|\rho(2T-t)\rangle\rangle dt,\label{eq:S3.1}
\end{equation}
where $|\rho(t)\rangle\rangle=U_{L}(t)|\rho_{0}\rangle\rangle$. The Hamiltonian deviation
$\delta H_{L}(t)$ in Liouville space is related to a Hamiltonian
deviation $\delta H(t)$ in Hilbert space. That is, $\delta H_{L}(t)=\delta H(t)\otimes I-I\otimes\delta H(t)^{T}$,
in such a way that the total Hamiltonian in Hilbert space reads $H(t)+\delta H(t)$.
This implies that in Hilbert space the vector $\delta H_{L}(t)|\rho(t)\rangle\rangle$
takes the form $[\delta H(t),\rho(t)]$ or, equivalently, 
\begin{equation}
\delta H_{L}(t)|\rho(t)\rangle\rangle=|[\delta H(t),\rho(t)]\rangle\rangle,\label{eq:S3.2}
\end{equation}
where $|A\rangle\rangle$ denotes a column vector associated with an arbitrary
matrix $A$ of dimension $n\times n$. 

Accordingly, we can write $\langle\langle\rho(t)|\delta H_{L}(t)|\rho(t)\rangle\rangle$
as 

\begin{equation}
\langle\langle\rho(t)|\delta H_{L}(t)|\rho(t)\rangle\rangle=\langle\langle\rho(t)|[\delta H(t),\rho(t)]\rangle\rangle.\label{eq:S3.3}
\end{equation}
By applying Eq. (\ref{eq:4 expect value in Liouville space}) to the
right hand side of (\ref{eq:S3.3}), it follows that 

\begin{align}
\langle\langle\rho(t)|\delta H_{L}(t)|\rho(t)\rangle\rangle & =\textrm{Tr}\left(\rho^{\dagger}(t)[\delta H(t),\rho(t)]\right)\nonumber \\
 & =\textrm{Tr}\left(\rho(t)\delta H(t)\rho(t)-\rho(t)\rho(t)\delta H(t)\right)\nonumber \\
 & =0,\label{eq:S3.4}
\end{align}
where in the second we apply the hermiticity of $\rho(t)$, and the
cyclic property of the trace is used in the third line. Since this
is equally valid if we replace $\rho(t)$ by any hermitian operator,
we conclude that $\langle\langle\rho_{0}|\Theta|\rho_{0}\rangle\rangle=0$.

\section*{Appendix III - Robustness to SPAM errors }

Here, we characterize SPAM errors and show that the estimation of
$\sigma_{n}$ is robust to leading corrections associated with these
errors. We start by deriving Eqs. (\ref{eq:34 Kp}) and (\ref{eq:35 Km}).
Equation (\ref{eq:34 Kp}) follows by applying once again the Magnus
expansion to $d_{t}|\rho_{int}\rangle\rangle=\Lambda_{\textrm{p}}(t)|\rho_{int}\rangle\rangle$,
where $\Lambda_{\textrm{p}}(t)$ contains the dissipator and the Hamiltonian
deviation (coherent errors) that affect the implementation of $U_{\textrm{p}}$,
written in the interaction picture. Moreover, $|\rho_{int}\rangle\rangle$
is defined similarly to Eq. (\ref{eq:S1.2}), with $U_{L}(t)$ replaced
by the (error-free) preparation unitary evaluated at time $t$. The
initial erroneous state is thus given by $U_{\textrm{p}}e^{\Omega_{\textrm{p}}}|0\rangle\rangle$,
being $\Omega_{\textrm{p}}=\sum_{i=1}^{\infty}\Omega_{\textrm{p},i}$
the Magnus expansion that accounts for preparation errors. In particular,
\begin{align}
\Omega_{\textrm{p},1} & =\int_{0}^{T_{\textrm{p}}}dt\Lambda_{\textrm{p}}(t),\label{eq:S4.1}\\
\Omega_{\textrm{p},2} & =\frac{1}{2}\int_{0}^{T_{\textrm{p}}}dt'\int_{0}^{t'}[\Lambda_{\textrm{p}}(t'),\Lambda_{\textrm{p}}(t)]dt,\label{eq:S4.2}
\end{align}
where $T_{\textrm{p}}$ is the time invested in the preparation stage. 

For the measurement stage, the target circuit is $U_{\textrm{p}}^{\dagger}$.
The error-prone dynamics yields an imperfect implementation $K_{\textrm{m}}=U_{\textrm{p}}^{\dagger}e^{\Omega_{\textrm{m}}}$
of $U_{\textrm{p}}^{\dagger}$, with a Magnus expansion $\Omega_{\textrm{m}}$
that in general differs from $\Omega_{\textrm{p}}$. In contrast with
Eq. (\ref{eq:35 Km}), in the previous expression for $K_{\textrm{m}}$
the ideal unitary $U_{\textrm{p}}^{\dagger}$ appears at the l.h.s.
of $e^{\Omega_{\textrm{m}}}$. The first step to address this issue
is to realize that we can express $K_{\textrm{m}}$ as $K_{\textrm{m}}=\left(U_{\textrm{p}}^{\dagger}e^{\Omega_{\textrm{m}}}U_{\textrm{p}}\right)U_{\textrm{p}}^{\dagger}$.
Since $\left(U_{\textrm{p}}^{\dagger}\Omega_{\textrm{m}}U_{\textrm{p}}\right)^{n}=U_{\textrm{p}}^{\dagger}\Omega_{\textrm{m}}^{n}U_{\textrm{p}}$
for any positive integer $n$, the unitaries $U_{\textrm{p}}^{\dagger}$
and $U_{\textrm{p}}$ can be absorbed into the the exponent of $K_{\textrm{m}}$,
i.e. $U_{\textrm{p}}^{\dagger}e^{\Omega_{\textrm{m}}}U_{\textrm{p}}=e^{U_{\textrm{p}}^{\dagger}\Omega_{\textrm{m}}U_{\textrm{p}}}:=e^{\bar{\Omega}_{\textrm{m}}}$.
Therefore, $K_{\textrm{m}}=e^{\bar{\Omega}_{\textrm{m}}}U_{\textrm{p}}^{\dagger}$,
in agreement with Eq. (\ref{eq:35 Km}). 

Now, let us characterize the corrections to $\sigma_{n}$ that result
from SPAM errors. We recall that these errors impact the measurement
of the quantities $R_{k}=\langle\langle\rho_{0}|e^{k\chi}|\rho_{0}\rangle\rangle=\langle\langle0|U_{\textrm{p}}^{\dagger}e^{k\chi}U_{\textrm{p}}|0\rangle\rangle$,
due to the non-ideal implementations of $U_{\textrm{p}}$ and $U_{\textrm{p}}^{\dagger}$.
Thus, instead of $R_{k}$ what is measured in practice is $R_{k}^{\prime}=\langle\langle\rho_{\textrm{m}}|e^{k\chi}|\rho_{\textrm{p}}\rangle\rangle$,
where $|\rho_{\textrm{p}}\rangle\rangle=U_{\textrm{p}}e^{\Omega_{\textrm{p}}}|0\rangle\rangle$
and $\langle\langle\rho_{\textrm{m}}|=\langle\langle0|e^{\bar{\Omega}_{\textrm{m}}}U_{\textrm{p}}^{\dagger}$.
This also substitutes the SPAM-free quantity $\sigma_{n}=\sum_{k=0}^{n}a_{k}^{(n)}R_{k}$
by $\sigma_{n}^{\prime}=\sum_{k=0}^{n}a_{k}^{(n)}R_{k}^{\prime}$. 

Our goal is to show that SPAM errors that dominate each term $R_{k}^{\prime}$
are suppressed in the calculation of $\sigma_{n}^{\prime}$. As we
shall see, the key property behind this suppression is the condition
(cf. Eq. (\ref{eq:25 ak eq}) for $j=0$)
\begin{equation}
\sum_{k=0}^{n}a_{k}^{(n)}=0.\label{eq:S4.3}
\end{equation}
By applying Eq. (\ref{eq:S4.3}), we have that 
\begin{align}
\sigma_{n}^{\prime} & =\sum_{k=1}^{n}a_{k}^{(n)}\left\{ R_{k}^{\prime}-R_{0}^{\prime}\right\} \nonumber \\
 & =\sum_{k=1}^{n}a_{k}^{(n)}\left\{ \langle\langle\rho_{\textrm{m}}|e^{k\chi}|\rho_{\textrm{p}}\rangle\rangle-\langle\langle\rho_{\textrm{m}}|\rho_{\textrm{p}}\rangle\rangle\right\} \nonumber \\
 & =\sum_{k=1}^{n}a_{k}^{(n)}\langle\langle0|\left\{ e^{\bar{\Omega}_{\textrm{m}}}U_{\textrm{p}}^{\dagger}e^{k\chi}U_{\textrm{p}}e^{\Omega_{\textrm{p}}}-e^{\bar{\Omega}_{\textrm{m}}}e^{\Omega_{\textrm{p}}}\right\} |0\rangle\rangle.\label{eq:S4.4}
\end{align}
In addition, for each $0\leq k\leq n$ we can expand the exponentials
in the last line of Eq. (\ref{eq:S4.4}) to obtain 
\begin{align}
e^{\bar{\Omega}_{\textrm{m}}}U_{\textrm{p}}^{\dagger}e^{k\chi}U_{\textrm{p}}e^{\Omega_{\textrm{p}}}-e^{\bar{\Omega}_{\textrm{m}}}e^{\Omega_{\textrm{p}}} & =e^{\bar{\Omega}_{\textrm{m}}}U_{\textrm{p}}^{\dagger}\sum_{l=1}^{\infty}\left(k\chi\right)^{l}U_{\textrm{p}}e^{\Omega_{\textrm{p}}}\nonumber \\
 & =U_{\textrm{p}}^{\dagger}\sum_{l=1}^{\infty}\left(k\chi\right)^{l}U_{\textrm{p}}+\sum_{l=1}^{\infty}k^{l}\bar{\chi}^{l}\sum_{l=1}^{\infty}\Omega_{\textrm{p}}^{l}\nonumber \\
 & \quad+\sum_{l=1}^{\infty}\bar{\Omega}_{\textrm{m}}^{l}\sum_{l=1}^{\infty}k^{l}\bar{\chi}^{l}+\sum_{l=1}^{\infty}\bar{\Omega}_{\textrm{m}}^{l}\sum_{l=1}^{\infty}k^{l}\bar{\chi}^{l}\sum_{l=1}^{\infty}\Omega_{\textrm{p}}^{l},\label{eq:S4.5}
\end{align}
where $\bar{\chi}:=U_{\textrm{p}}^{\dagger}\chi U_{\textrm{p}}$.

The first term in the second line of Eq. (\ref{eq:S4.5}) can be rewritten
as $U_{\textrm{p}}^{\dagger}e^{k\chi}U_{\textrm{p}}-I_{L}$. Therefore,
its contribution to $\sigma_{n}^{\prime}$ is given by 
\begin{align}
\sum_{k=1}^{n}a_{k}^{(n)}\langle\langle0|\left\{ U_{\textrm{p}}^{\dagger}e^{k\chi}U_{\textrm{p}}-I_{L}\right\} |0\rangle\rangle & =\sum_{k=1}^{n}a_{k}^{(n)}\left\{ \langle\langle\rho_{0}|e^{k\chi}|\rho_{0}\rangle\rangle-\langle\langle\rho_{0}|\rho_{0}\rangle\rangle\right\} \nonumber \\
 & =\sigma_{n},\label{eq:S4.6}
\end{align}
while the remaining terms provide corrections to $\sigma_{n}$. According
to Eq. (\ref{eq:S4.5}), the leading-order terms in the difference
$\sigma_{n}^{\prime}-\sigma_{n}$ are $\langle\langle0|\bar{\chi}\Omega_{\textrm{p}}|0\rangle\rangle$
and $\langle\langle0|\bar{\Omega}_{\textrm{m}}\bar{\chi}|0\rangle\rangle$, which
constitute quadratic corrections to $\sigma_{n}$. Hence, $\sigma_{n}^{\prime}$
is robust to the leading (linear) corrections that appear in each
$R_{k}^{\prime}$. For example, $R_{1}^{\prime}$ is affected by the
linear error terms $\langle\langle0|\bar{\chi}|0\rangle\rangle$, $\langle\langle0|\Omega_{\textrm{p}}|0\rangle\rangle$,
and $\langle\langle0|\bar{\Omega}_{\textrm{m}}|0\rangle\rangle$. We also remark
that the SPAM contributions $\langle\langle0|\sum_{l=1}^{\infty}\Omega_{\textrm{p}}^{l}|0\rangle\rangle$,
$\langle\langle0|\sum_{l=1}^{\infty}\bar{\Omega}_{\textrm{m}}^{l}|0\rangle\rangle$,
and $\langle\langle0|\sum_{l=1}^{\infty}\Omega_{\textrm{p}}^{l}\sum_{l=1}^{\infty}\bar{\Omega}_{\textrm{m}}^{l}|0\rangle\rangle$,
characteristic of each $R_{k}^{\prime}$, are absent in $\sigma_{n}^{\prime}$. 

Taking into account Eqs. (\ref{eq:S4.5}) and (\ref{eq:S4.6}), the
application of the triangular inequality yields 

\begin{align}
|\sigma_{n}^{\prime}-\sigma_{n}| & \leq\sum_{k=1}^{n}\left|a_{k}^{(n)}\right|\left\{ \sum_{l,l'=1}^{\infty}k^{l}\left|\langle\langle0|\bar{\chi}^{l}\Omega_{\textrm{p}}^{l'}|0\rangle\rangle\right|+\sum_{l,l'=1}^{\infty}k^{l}\left|\langle\langle0|\bar{\Omega}_{\textrm{m}}^{l'}\bar{\chi}^{l}|0\rangle\rangle\right|\right.\nonumber \\
 & \quad+\left.\sum_{l,l',l''=1}^{\infty}k^{l}\left|\langle\langle0|\bar{\Omega}_{\textrm{m}}^{l'}\bar{\chi}^{l}\Omega_{\textrm{p}}^{l''}|0\rangle\rangle\right|\right\} .\label{eq:S4.7}
\end{align}
In addition, the definition of the spectral norm $\left\Vert \ast\right\Vert $
and the fact that $\langle\langle0|0\rangle\rangle=1$ allows us to bound each term
in Eq. (\ref{eq:S4.7}). For example, for the leading order terms
we have that 

\begin{align}
|\langle\langle0|\bar{\chi}\Omega_{\textrm{p}}|0\rangle\rangle| & \leq\left\Vert \bar{\chi}\Omega_{\textrm{p}}\right\Vert \leq\left\Vert \chi\right\Vert \left\Vert \Omega_{\textrm{p}}\right\Vert ,\label{eq:S4.9}\\
|\langle\langle0|\bar{\Omega}_{\textrm{m}}\bar{\chi}|0\rangle\rangle| & \leq\left\Vert \bar{\Omega}_{\textrm{m}}\bar{\chi}\right\Vert \leq\left\Vert \Omega_{\textrm{m}}\right\Vert \left\Vert \chi\right\Vert ,\label{eq:S4.10}
\end{align}
where the definition of $\left\Vert \ast\right\Vert $ is used in
the middle inequalities, and the rightmost inequalities follow from
the submultiplicativity of $\left\Vert \ast\right\Vert $ and its
unitary invariance. From Eq. (\ref{eq:S4.7}), it is also clear that
a generic term at the r.h.s. has an upper bound proportional to $\left\Vert \Omega_{\textrm{m}}\right\Vert ^{l'}\left\Vert \chi\right\Vert ^{l}\left\Vert \Omega_{\textrm{p}}\right\Vert ^{l''}$,
with $l'+l''\geq1$ and $l\geq1$. 

\section*{Appendix IV - Mitigation of local readout errors}

Readout errors lead to incorrect measurements in quantum computing.
Usually, these measurements are projective measurements in the computational
basis, and determine the state of each qubit as being ``0'' or ``1''.
A readout error occurs when the true state is 0 and it is registered
as 1 or vice versa. In a system composed of $N$ qubits, it is possible
to characterize readout errors by using a $2^{N}\times2^{N}$ ``detector
matrix'' $\mathbf{D}$. An element $\mathbf{D}_{i,j}$ of this matrix
represents the probability of measuring the ($N$-qubit) computational
state $i$, given that the true state is $j$.

Suppose now that we perform a measurement after implementing some
generic quantum circuit. If $q_{i}$ denotes the probability of measuring
the computational state $i$, in the presence of readout errors, the
probability vector $\boldsymbol{q}=(q_{1},q_{2},...q_{2^{N}})^{\textrm{T}}$
can be written as 
\begin{equation}
\boldsymbol{q}=\mathbf{D}\boldsymbol{p},\label{eq:S5.1 detector matrix}
\end{equation}
where $\boldsymbol{p}=(p_{1},p_{2},...p_{2^{N}})^{\textrm{T}}$ and
$p_{i}$ is the probability for measuring $i$ in the absence of readout
errors. Thus, the ``error-free'' (note that the probabilities $p_{i}$
can include preparation errors and circuit errors but by definition
they exclude readout errors) probabilities $\boldsymbol{p}$ are given
by

\begin{equation}
\boldsymbol{p}=\mathbf{D}^{-1}\boldsymbol{q}.\label{eq:S5.2 inverse of detector matrix}
\end{equation}

A column $\{\mathbf{D}_{i,j}\}_{i}$ of the detector matrix $\mathbf{D}$
can be experimentally determined by preparing the computational state
$j$ and measuring the corresponding probability distribution. If
$N$ is small, the full matrix $\mathbf{D}$ can be constructed from
the measurements performed on all the computational states. However,
such a procedure is not scalable because there are $2^{N}$ computational
states. On the other hand, one can model the readout errors by assuming
that they contain negligible correlations. If this assumption is well
founded, the matrix $\mathbf{D}$ takes the form of a tensor product 

\begin{equation}
\mathbf{D}=\mathbf{D}_{1}\otimes\mathbf{D}_{2}\otimes...\mathbf{D}_{k}\otimes...,\label{eq:S5.3 local readout errors}
\end{equation}
where the matrices $\mathbf{D}_{k}$ represent local detector matrices
acting on subsets of qubits. In this case, it is possible to efficiently
determine the matrix $\mathbf{D}$, if all the matrices $\mathbf{D}_{k}$
have small dimension and this dimension is independent of $N$. For
example, completely uncorrelated readout errors would be characterized
by measuring $N$ single-qubit detector matrices. The mitigation of
local readout errors is performed through the inverse 
\begin{equation}
\mathbf{D}^{-1}=\mathbf{D}_{1}^{-1}\otimes\mathbf{D}_{2}^{-1}\otimes...\mathbf{D}_{k}^{-1}\otimes...,\label{eq:S5.4 inverse of local detector matrix}
\end{equation}
which can also be efficiently computed for local detector matrices
of small dimension. 

For completely uncorrelated readout errors, preparing
the two computational states where all the qubits are in the ground
(``0'') state or all the qubits are in the excited (``1'') state
is sufficient to evaluate $\mathbf{D}$. This procedure is equivalent
to a simultaneous preparation of all the qubits in any of its two
computational states. Thus, the associated readout statistics provides
all the information required to determine the local detector matrices
$\mathbf{D}_{k}$.

\section*{Appendix V - Description of error sources in numerical simulations }

In this appendix, we provide a detailed description of the error models
used for the simulations of Sec. \textcolor{blue}{V}. First, we will
characterize the error sources associated with the simulation of the
GHZ state, and then the error sources simulated in the calculation
of the average incoherent infidelity. In both cases the CNOT gates
are based on the cross resonance interaction (cf. Eq. (\ref{eq:37.2 cross reson CNOT})). 

\subsection*{Error sources for the simulation of the GHZ state }

We assume that two-qubit gates are the only gates affected by noise.
More specifically, we consider local dephasing and amplitude damping
that act alongside the cross resonance pulse used for the implementation
of CNOT gates. In the case of the circuit used for preparing the five-qubit
GHZ state, these noise channels operate on all the qubits of the quantum
register and not only on the specific pair that undergoes the cross
resonance interaction. The corresponding error-free evolution is given
by 
\begin{equation}
U_{GHZ}=U_{CNOT}^{(4,5)}U_{CNOT}^{(3,4)}U_{CNOT}^{(2,3)}U_{CNOT}^{(1,2)}U_{had}^{(1)},\label{eq:37.3 ideal GHZ unitary-1}
\end{equation}
where $U_{had}^{(1)}$ represents a Hadamard gate acting on qubit
1. We also remark that in Eq. (\ref{eq:37.3 ideal GHZ unitary-1})
we use the standard notation for unitary operators that act on density
matrices. In particular, the associated CNOT gates obey Eq. (\ref{eq:37.2 cross reson CNOT}). 

The effect of noise is introduced through the superoperators 
\begin{align}
\hat{D}[\rho] & =\sum_{i=1}^{5}\left\{ Z_{i}\rho Z_{i}^{\dagger}-\frac{1}{2}Z_{i}^{\dagger}Z_{i}\rho-\frac{1}{2}\rho Z_{i}^{\dagger}Z_{i}\right\} ,\label{eq:37.3 dephasing in HS}\\
\hat{A}[\rho] & =\sum_{i=1}^{5}\left\{ \sigma_{-}^{(i)}\rho\sigma_{+}^{(i)}-\frac{1}{2}\sigma_{+}^{(i)}\sigma_{-}^{(i)}\rho-\frac{1}{2}\rho\sigma_{+}^{(i)}\sigma_{-}^{(i)}\right\} ,\label{eq:37.4 ampl damp in HS}
\end{align}
where $\hat{D}$ stands for dephasing, $\hat{A}$ stands for amplitude
damping, and $\sigma_{-}^{(i)}$ is the annihilation operator $\left(\begin{array}{cc}
0 & 1\\
0 & 0
\end{array}\right)$ acting on the qubit $i$ (in addition, $\sigma_{+}^{(i)}=\sigma_{-}^{(i)\dagger}$).\textcolor{blue}{{}
}By adding Eqs. (\ref{eq:37.3 dephasing in HS}) and (\ref{eq:37.4 ampl damp in HS})
to the unitary dynamics generated by the cross resonance Hamiltonian
$Z_{j}\otimes X_{k}$ (which includes also the contribution from coherent
errors), we obtain a\textcolor{blue}{{} }GKLS (Gorini-Kossakowski-Sudarshan-Lindblad)
master equation for the error-prone evolution. We consider equal relaxation
rates for $\hat{D}$ and $\hat{A}$, which leads to the total noise
superoperator 
\begin{equation}
\hat{L}[\rho]=\xi\left\{ \frac{1}{2}\hat{D}[\rho]+\frac{1}{2}\hat{A}[\rho]\right\} .\label{eq:37.5 dissipator in HS}
\end{equation}
The parameter $\xi$ characterizes the strength of the noise. Moreover,
$\hat{L}$ generates the dissipator $\mathcal{L}$ in the Liouville
space formalism, as described below. 

For the Liouville space representation of the noise channels $\hat{D}$
and $\hat{A}$, we use the calligraphic letters $\mathcal{D}$ and
$\mathcal{A}$, respectively. In this representation, such superoperators
become $N^{2}\times N^{2}=2^{10}\times2^{10}$ matrices given by \cite{gyamfi2020fundamentals}

\begin{align}
\mathcal{D} & =\sum_{i=1}^{5}\left\{ Z_{i}\otimes Z_{i}-I_{L}\right\} ,\label{eq:37.6 dephasing in LS}\\
\mathcal{A} & =\sum_{i=1}^{5}\left\{ \sigma_{i}^{(-)}\otimes\sigma_{i}^{(-)}-\frac{1}{2}\sigma_{i}^{(+)}\sigma_{i}^{(-)}\otimes I-\frac{1}{2}I\otimes\sigma_{i}^{(+)}\sigma_{i}^{(-)}\right\} .\label{eq:37.7 Ampl damp in LS}
\end{align}
Importantly, each matrix labeled by $i$ in Eqs. (\ref{eq:37.6 dephasing in LS})
and (\ref{eq:37.7 Ampl damp in LS}) is an $N\times N$ matrix that
includes an implicit tensor product with a $2^{4}\times2^{4}$ identity
matrix, acting on all the qubits different from $i$. The corresponding
dissipator reads 
\begin{equation}
\mathcal{L}=\xi\left\{ \frac{1}{2}\mathcal{D}+\frac{1}{2}\mathcal{A}\right\} .\label{eq:37.9 dissipator in LS}
\end{equation}
Since the Hamiltonian $Z_{j}\otimes X_{k}$ is time independent, and
noise is only active during the cross resonance pulse, $\mathcal{L}$
retains also its time independence. 

Therefore, the error-prone implementation of $U_{CNOT}^{(j,k)}$ is
governed by the master equation 
\begin{align}
d_{t}\ket{\rho} & =\left(-iH_{L}^{(j,k)}-i\delta H_{L}^{(j,k)}+\mc L\right)\ket{\rho},\label{eq:37.11 master eq for CNOT}\\
H_{L}^{(j,k)} & =\left(Z_{j}\otimes X_{k}\right)\otimes I-I\otimes\left(Z_{j}\otimes X_{k}\right)^{\textrm{T}},\label{eq:37.12 LS Ham for CNOT}\\
\delta H_{L}^{(j,k)} & =\eta\left\{ \left(Z_{j}\otimes Z_{k}\right)\otimes I-I\otimes\left(Z_{j}\otimes Z_{k}\right)^{\textrm{T}}\right\} ,\label{eq:37.13 ZZ term in LS}
\end{align}
where $\mc L$ is given by Eq. (\ref{eq:37.9 dissipator in LS}).
In Eqs. (\ref{eq:37.12 LS Ham for CNOT}) and (\ref{eq:37.13 ZZ term in LS}),
it is important to stress that the terms $Z_{j}\otimes X_{k}$ and
$Z_{j}\otimes Z_{k}$ are $N\times N$ matrices that act as the identity
for all the qubits different from $j$ and $k$. Equation (\ref{eq:37.11 master eq for CNOT})
can be directly integrated to obtain the exact solution $e^{\left(-iH_{L}^{(j,k)}T-i\delta H_{L}^{(j,k)}+\mathcal{L}\right)T}$,
where $T$ is the time invested in the cross resonance interaction.

In this simulation, we assume that all the single-qubit gates are
error-free. Accordingly, the error-prone implementation of $U_{CNOT}^{(j,k)}$
is given by 
\begin{align}
K_{CNOT}^{(j,k)} & =e^{i\frac{\pi}{4}X_{k}}\otimes e^{-i\frac{\pi}{4}X_{k}}K_{CR}^{(j,k)}e^{i\frac{\pi}{4}Z_{j}}\otimes e^{-i\frac{\pi}{4}Z_{j}},\label{eq:37.14 KCNOT}\\
K_{CR}^{(j,k)} & =e^{\left(-iH_{L}^{(j,k)}-i\delta H_{L}^{(j,k)}+\mathcal{L}\right)T},\label{eq:37.15 noisy CR}
\end{align}
where the rotations $e^{i\frac{\pi}{4}X_{k}}$ and $e^{i\frac{\pi}{4}Z_{j}}$
are expressed using the representation of unitary operators in Liouville
space. Specifically, the matrix corresponding to a unitary $U$ reads
$U\otimes U^{\ast}$, where $U^{\ast}$ is the elementwise complex
conjugate of $U$. Since $U_{had}^{(1)}$ is a real matrix, the error-prone
implementation of $U_{GHZ}$ is given by 
\begin{equation}
K=K_{CNOT}^{(4,5)}K_{CNOT}^{(3,4)}K_{CNOT}^{(2,3)}K_{CNOT}^{(1,2)}U_{had}^{(1)}\otimes U_{had}^{(1)}.\label{eq:37.16 K for GHZ circuit}
\end{equation}

Let us now characterize the inverse evolution $K_{I}$, following
the pulse inverse approach on which our theory is based. Due to the
time independence of $Z_{j}\otimes X_{k}$, the driving that reverses
the corresponding CNOT in the error-free scenario is simply $-Z_{j}\otimes X_{k}$.
Thus, the ideal inverse for this gate reads 

\begin{equation}
U_{I,CNOT}^{(j,k)}=e^{-i\frac{\pi}{4}Z_{j}}U_{CR}^{\dagger}e^{-i\frac{\pi}{4}X_{k}}.\label{eq:37.17 inverse CNOT}
\end{equation}
Since the dissipator remains invariant under a sign change of the
Hamiltonian, in the error-prone execution of $U_{I,CNOT}^{(j,k)}$
noise is also characterized by Eq. (\ref{eq:37.9 dissipator in LS}).
Furthermore, the parasitic term $Z_{j}\otimes Z_{k}$ is also unaffected
by this sign reversal \cite{krinner2020benchmarking}. Accordingly, the
implementation of $U_{I,CNOT}^{(j,k)}$ is given by 
\begin{equation}
K_{I,CNOT}^{(j,k)}=e^{-i\frac{\pi}{4}Z_{j}}\otimes e^{i\frac{\pi}{4}Z_{j}}K_{I,CR}^{(j,k)}e^{-i\frac{\pi}{4}X_{k}}\otimes e^{i\frac{\pi}{4}X_{k}},\label{eq:37.18 KI,CNOT}
\end{equation}
where
\begin{equation}
K_{I,CR}^{(j,k)}=e^{\left(iH_{L}^{(j,k)}-i\delta H_{L}^{(j,k)}+\mathcal{L}\right)T}.\label{eq:37.19 KI,CR}
\end{equation}
This leads to the inverse evolution 

\begin{equation}
K_{I}=U_{had}^{(1)}\otimes U_{had}^{(1)}K_{I,CNOT}^{(1,2)}K_{I,CNOT}^{(2,3)}K_{I,CNOT}^{(3,4)}K_{I,CNOT}^{(4,5)}.\label{eq:37.20 KI for GHZ circuit}
\end{equation}

For the sake of clarity, we remark that the exact infidelity $\varepsilon_{\textrm{inc}}(\varrho^{(\textrm{id})},\tilde{\varrho})$
is computed with the final states 
\begin{align}
|\varrho^{(\textrm{id})}\rangle\rangle & =U_{GHZ}\otimes U_{GHZ}^{\ast}|\rho_{0}\rangle\rangle,\label{eq:37.21 final error-free GHZ state}\\
|\tilde{\varrho}\rangle\rangle & =\tilde{K}|\rho_{0}\rangle\rangle,\label{eq:37.22 noisy GHZ state}
\end{align}
where $\tilde{K}$ denotes the implementation of $U_{GHZ}$ that excludes
coherent errors. Specifically, 
\begin{equation}
\tilde{K}=\tilde{K}_{CNOT}^{(4,5)}\tilde{K}_{CNOT}^{(3,4)}\tilde{K}_{CNOT}^{(2,3)}\tilde{K}_{CNOT}^{(1,2)}U_{had}^{(1)}\otimes U_{had}^{(1)},\label{eq:K without coherent errors}
\end{equation}
where $\tilde{K}_{CNOT}^{(j,k)}=e^{i\frac{\pi}{4}X_{k}}\otimes e^{-i\frac{\pi}{4}X_{k}}\left[e^{\left(-iH_{L}^{(j,k)}+\mathcal{L}\right)T}\right]e^{i\frac{\pi}{4}Z_{j}}\otimes e^{-i\frac{\pi}{4}Z_{j}}$.
Furthermore, the estimates of $\varepsilon_{\textrm{inc}}(\varrho^{(\textrm{id})},\tilde{\varrho})$
through $\sigma_{n}$ are evaluated with the evolutions $K$ and $K_{I}$
in Eqs. (\ref{eq:37.16 K for GHZ circuit}) and (\ref{eq:37.20 KI for GHZ circuit}). 

\subsection*{Error sources for the simulation of the average incoherent infidelity}

In this example, we simulate the average incoherent infidelity of
a CNOT gate, with respect to input states prepared by two-qubit Clifford
gates. Any two-qubit Clifford gate can be obtained by combining single-qubit
Clifford gates with the CNOT gate, the SWAP gate, and the iSWAP gate (see Supplementary Information of \cite{barends2014superconducting}). Moreover, it is possible to implement
the SWAP gate and the iSWAP gate using CNOT gates and single-qubit
Clifford gates \cite{barends2014superconducting}.
This implies that any two-qubit Clifford gate can be compiled using
single-qubit Clifford gates and CNOT gates. Thus, our simulation of
errors comprises noise and the $Z_{1}\otimes Z_{2}$ coherent error,
in the case of CNOT gates, and the single-qubit coherent errors described
in Eqs. (\ref{eq:37.26 single-qub coh error 1}) and (\ref{eq:37.27 single-qub coh error 2}).
These error sources affect both the target CNOT and the Clifford gates
used for preparation and measurement. As in the previous example,
we also introduce noise as local dephasing and amplitude damping acting
alongside the cross resonance interaction. However, now we consider
the dissipator 
\begin{equation}
\mathcal{L}=\xi\left\{ \mathcal{D}+\frac{1}{10}\mathcal{A}\right\} ,\label{eq:37.28 dissipator for Cliffords}
\end{equation}
which gives more weight to the dephasing channel $\mathcal{D}$. We
also remark that for the channels $\mathcal{D}$ and $\mathcal{A}$
in Eq. (\ref{eq:37.28 dissipator for Cliffords}) the summation limit
in (\ref{eq:37.6 dephasing in LS}) and (\ref{eq:37.7 Ampl damp in LS})
must be replaced by $2$. 

In addition to the aforementioned error sources, we include readout
errors and errors in the preparation of the fiducial state $|0\rangle\rangle$.
Based on Reference \cite{landa2022experimental},
we consider a rotated fiducial state $|\hat{0}\rangle\rangle$, obtained
by applying the rotation $R_{Y}(0.005\pi)R_{X}(0.005\pi)$ to the
``0'' state of each qubit. In this way, an initial state $|\rho_{0}^{(m)}\rangle\rangle$
is the result of applying an error-prone Clifford gate $K_{\textrm{p}}^{(m)}$
to $|\hat{0}\rangle\rangle$. That is, 
\begin{equation}
|\rho_{0}^{(m)}\rangle\rangle=K_{\textrm{p}}^{(m)}|\hat{0}\rangle\rangle,\label{eq:37.29 initial state with fiducial error}
\end{equation}
where the superscript $m$ labels a gate randomly chosen from the
set of two-qubit Clifford gates. The infidelities $\bar{\varepsilon}_{\textrm{inc}}$
and $\bar{\varepsilon}$ are thus computed using the final states
$|\varrho^{(m)}\rangle\rangle=\tilde{K}|\rho_{0}^{(m)}\rangle\rangle$ and $|\varrho^{(m)}\rangle\rangle=K|\rho_{0}^{(m)}\rangle\rangle$,
with $|\rho_{0}^{(m)}\rangle\rangle$ given in Eq. (\ref{eq:37.29 initial state with fiducial error}).

On the other hand, we incorporate single-qubit readout errors by using
a noisy POVM (positive operator valued measurement) to characterize
the corresponding measurement. This POVM has elements $\Pi$ and $I_{2\times2}-\Pi$,
where $I_{2\times2}=\left(\begin{array}{cc}
1 & 0\\
0 & 1
\end{array}\right)$ and \cite{landa2022experimental}
\begin{equation}
\Pi=\sum_{k}\pi_{k}P_{k},\quad P_{k}\in\left\{ I_{2\times2},X,Y,Z\right\} .\label{eq:37.30 POVM element}
\end{equation}
In the absence of readout errors, $\Pi=\frac{1}{2}\left(I_{2\times2}+Z\right):=\Pi^{(\textrm{id})}$
is the projector on the ``0'' state. This corresponds to $\pi_{0}=\pi_{3}=\frac{1}{2}$.
For the noisy POVM, we consider the coefficients 
\begin{align}
\pi_{0} & =0.501,\label{eq:37.31 pi_0}\\
\pi_{3} & =0.495,\label{eq:37.32 pi_3}\\
\pi_{1} & =\pi_{2}=0.\label{eq:37.33 pi_1}
\end{align}
In this way, the ideal row vector $\langle\langle0|=\left\langle\langle   \Pi_{1}^{(\textrm{id})}\otimes\Pi_{2}^{(\textrm{id})}\right|$
is substituted by $\left\langle\langle \Pi_{12}\right|:=\left\langle\langle \Pi_{1}\otimes\Pi_{2}\right|$,
where the subindices 1,2 label each qubit and both $\Pi_{1}$ and
$\Pi_{2}$ are characterized by the coefficients (\ref{eq:37.31 pi_0})-(\ref{eq:37.33 pi_1}).
We note that the resulting POVM describes a small error affecting
measurements in the computational basis. Specifically, the local detector
matrix (see Appendix \textcolor{blue}{IV}) is given by 
\begin{align}
\mathbf{D}_{k} & =\left(\begin{array}{cc}
\pi_{0}+\pi_{3} & \pi_{0}-\pi_{3}\\
1-(\pi_{0}+\pi_{3}) & 1-(\pi_{0}-\pi_{3})
\end{array}\right)\nonumber \\
 & =\left(\begin{array}{cc}
0.996 & 0.006\\
0.004 & 0.994
\end{array}\right),\label{eq:37.34 detect matrix with noisy POVM}
\end{align}
where $k=1,2$. 

For the $m$th Clifford gate, the $n$th-order estimation of the corresponding
incoherent infidelity $\varepsilon_{\textrm{inc}}(\varrho_{m}^{(\textrm{id})},\tilde{\varrho}_{m})$
is performed through the quantity 

\begin{equation}
\sigma_{n}^{\prime(m)}=\sum_{k=0}^{n}a_{k}^{(n)}\left\langle\langle \Pi_{12}\right|K_{\textrm{m}}^{(m)}\left(K_{I}K\right)^{k}K_{\textrm{p}}^{(m)}\left|\hat{0}\right\rangle\rangle ,\label{eq:37.35 sigma_n for mth Clifford}
\end{equation}
where $K_{\textrm{m}}^{(m)}$ is the error-prone Clifford gate used
for the measurement. Therefore, we estimate the average incoherent
infidelity using $-\frac{1}{2M}\sum_{m=1}^{M}\sigma_{n}^{\prime(m)}$.
The error-prone target evolution and its inverse are given by 
\begin{align}
K & =e^{i\frac{\pi}{4}X_{2}}e^{i\theta Z_{2}}K_{CR}^{(1,2)}e^{i\frac{\pi}{4}Z_{1}},\label{eq:37.36 error-prone cnot with single-qub coherent error}\\
K_{I} & =e^{-i\frac{\pi}{4}Z_{1}}K_{I,CR}^{(1,2)}e^{-i\theta Z_{2}}e^{-i\frac{\pi}{4}X_{2}},\label{eq:37.37 error-prone inv cnot with single-qub coherent error}
\end{align}
where $K_{CR}^{(1,2)}$ and $K_{I,CR}^{(1,2)}$ satisfy Eqs. (\ref{eq:37.15 noisy CR})
and (\ref{eq:37.19 KI,CR}), respectively. Moreover, $K_{\textrm{m}}^{(m)}=K_{I,\textrm{p}}^{(m)}$,
meaning that $K_{\textrm{m}}^{(m)}$ is obtained from $K_{\textrm{p}}^{(m)}$
by applying the inverse gates in reversed order. 

\section*{Appendix VI - Estimation of the infidelity using Interleaved RB}

To estimate the infidelity $\bar{\varepsilon}$ associated with the
CNOT gate, we follow Reference \cite{magesan2012efficient}.
The Interleaved RB protocol is based on the original RB protocol used
to estimate the average gate infidelity \cite{magesan2011scalable}.
This is done by implementing random sequences of Clifford gates, which
are followed by a final gate that inverts each sequence and generates
the identity operation in the error-free case, and by measuring the
probability to obtain the initial state $|0\rangle$. The sequence
length $l$ is the number of Clifford gates that compose it, including
the gate that performs the reversion. 

According to the RB theory \cite{magesan2011scalable}, for a sequence
of length $l$ it holds that 

\begin{equation}
F_{l}=A\alpha^{l}+B,\label{eq:37.38 fit for RB}
\end{equation}
where $F_{l}$ is the measured survival probability (probability of
measuring $|0\rangle$), and $\alpha,$ $A$ and $B$ are constants
to be experimentally determined. $A$ and $B$ account for SPAM errors,
and $\alpha$ is used to estimate of the average
gate infidelity. The average gate infidelity represents
the infidelity averaged over Clifford gates as well as over the pure
initial states. For $n$-qubit Clifford gates, this
quantity is estimated in RB by
\begin{equation}
r_{\textrm{ave}}=\frac{2^{n}-1}{2^{n}}(1-\alpha).\label{eq:37.39 average gate infidelity}
\end{equation}
The decay rate $\alpha$ is derived by fitting Eq. (\ref{eq:37.38 fit for RB})
to the experimental data points $(l,F_{l})$, measured for different
sequence lengths $1\leq l\leq L$. 

The Interleaved RB method combines this procedure with a similar protocol,
where the random Clifford gates in a given sequence are interleaved
with the (fixed) target gate. As in RB, the composition with the final
gate must produce the identity operation in the absence of errors.
Moreover, the sequence length is still defined as the number of random
Clifford gates. The behavior of the corresponding survival probability
can also be modeled using an exponential decay 
\begin{equation}
\bar{F}_{l}=A\bar{\alpha}^{l}+B,\label{eq:37.40 fit for IRB}
\end{equation}
where $\bar{F}_{l}$ is the survival probability for a sequence of
length $l$. Therefore, the decay rate $\bar{\alpha}$ is obtained
by fitting Eq. (\ref{eq:37.40 fit for IRB}) to the experimental data
$(l,\bar{F}_{l})$. Using the parameters $\alpha$ and $\bar{\alpha}$,
Interleaved RB yields the following estimate of the average infidelity
for the target gate:

\begin{equation}
r=\frac{2^{n}-1}{2^{n}}\left(1-\frac{\bar{\alpha}}{\alpha}\right).\label{eq:37.41 gate infidelity}
\end{equation}

In our simulation, we consider sequence lengths such that $l=3+15(k)$,
with $0\leq k\leq20$. For each value of $l$, $F_{l}$ and $\bar{F}_{l}$
are survival probabilities averaged over 60 random Clifford sequences.
We use the NonlinearModelFit function in the Mathematica software
to derive the parameters $\alpha$ and $\bar{\alpha}$. The corresponding
average infidelity $r$ is evaluated for each value of the angles
$\theta$ and $\phi$, to obtain the orange curves in Fig. \textcolor{blue}{2}. NonlinearModelFit
also allows us to compute errors $\Delta\alpha$ and $\Delta\bar{\alpha}$
associated with the quality of the fittings based on Eqs. (\ref{eq:37.38 fit for RB})
and (\ref{eq:37.40 fit for IRB}). Using error propagation, the error
bars for the orange curves are evaluated as 

\begin{equation}
\Delta r=\frac{2^{n}-1}{2^{n}}\sqrt{\frac{\bar{\alpha}^{2}(\Delta\alpha)^{2}+\alpha^{2}(\Delta\bar{\alpha})^{2}}{\alpha^{4}}}.\label{eq:37.42 error for estimate of gate infidelity}
\end{equation}

In Fig. \textcolor{blue}{3}, we show the RB and Interleaved RB simulations
corresponding to Eqs. (\ref{eq:37.38 fit for RB}) and (\ref{eq:37.40 fit for IRB}),
respectively. These simulations are performed for $\theta=0.05$,
$\phi=0.04$, and $\xi=0.001$. The red dots are the averages $F_{l}$
and $\bar{F}_{l}$ (taken over the data depicted by the gray dots),
and the dashed blue curves are the fitting curves. Importantly, all
the errors sources described in Appendix \textcolor{blue}{V} are included in these simulations.
Figure \textcolor{blue}{3} shows the good quality of the fits corresponding
to the aforementioned parameters, which we also corroborated with
the simulations using other error parameters. This evidences that,
in our example, the substantial disparity between the curves for $r$
and $\bar{\varepsilon}$ in Fig. \textcolor{blue}{2} is a consequence of the inability
of Interleaved RB to predict the actual average infidelity. 
\begin{figure}

\begin{centering}
\includegraphics[scale=0.5]{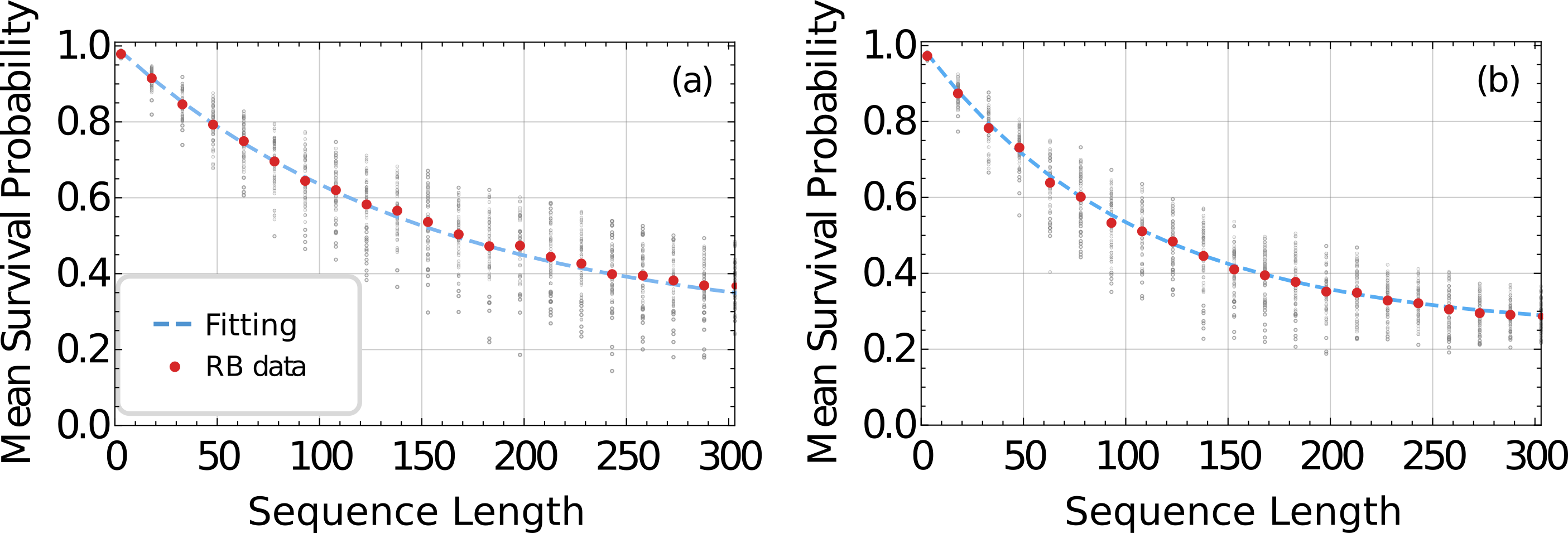}\caption{Interleaved RB simulations for the estimate of the (CNOT) gate infidelity
corresponding to $\theta=0.05$, $\phi=0.04$, and $\xi=0.001$. (a)
The RB data (red dots) for the survival probability averaged over
random sequences of Clifford gates. For each sequence length the average
is taken over 60 samples (gray dots). The dashed blue curve gives
the fit according to Eq. (\ref{eq:37.38 fit for RB}). (b) Interleaved
RB data (red dots) and fit (\ref{eq:37.40 fit for IRB}) (dashed blue
curve). In this case the 60 random sequences per each sequence length
contain Clifford gates interleaved with the CNOT gate. }
\par\end{centering}
\end{figure}

\section*{Appendix VII - Derivation of the bounds (\ref{eq:41 error bound 1})
and (\ref{eq:42 error bound 2}) }

To obtain the bounds (\ref{eq:41 error bound 1}) and (\ref{eq:42 error bound 2}),
we express the solutions to Eqs. (\ref{eq:S1.3}) and (\ref{eq:S1.11.1})
using the Dyson series. For Eq. (\ref{eq:S1.3}), we have that 
\begin{equation}
e^{\Omega(T)}=\sum_{k=0}^{\infty}\frac{1}{k!}\intop_{0}^{T}dt_{1}\intop_{0}^{T}dt_{2}...\intop_{0}^{T}dt_{k}\mathcal{T}\mathcal{L}^{int}(t_{1})\mathcal{L}^{int}(t_{2})...\mathcal{L}^{int}(t_{k}),\label{eq:S3.5}
\end{equation}
where $\mathcal{T}$ is the time ordering operator. The first and
second terms in the Dyson expansion (\ref{eq:S3.5}) are $I_{L}$
and $\Omega_{1}$, respectively. Therefore, 
\begin{align}
\left|\varepsilon_{\textrm{inc}}(\varrho,\tilde{\sigma})+\left\langle \Omega_{1}\right\rangle \right| & =\left|\langle\langle\rho_{0}|I_{L}-e^{\Omega(T)}+\Omega_{1}|\rho_{0}\rangle\rangle\right|\nonumber \\
 & \leq\left\Vert I_{L}-e^{\Omega(T)}+\Omega_{1}\right\Vert \nonumber \\
 & =\left\Vert \sum_{k=2}^{\infty}\frac{1}{k!}\intop_{0}^{T}dt_{1}\intop_{0}^{T}dt_{2}...\intop_{0}^{T}dt_{k}\mathcal{T}\mathcal{L}^{int}(t_{1})...\mathcal{L}^{int}(t_{k})\right\Vert \nonumber \\
 & \leq\sum_{k=2}^{\infty}\frac{1}{k!}\left(\intop_{0}^{T}dt\left\Vert \mathcal{L}^{int}(t)\right\Vert \right)^{k}\nonumber \\
 & =\sum_{k=2}^{\infty}\frac{1}{k!}\left(\intop_{0}^{T}dt\left\Vert \mathcal{L}(t)\right\Vert \right)^{k},\label{eq:S3.6}
\end{align}
where $\left\Vert \ast\right\Vert $ stands for the spectral norm.
The second line of Eq. (\ref{eq:S3.6}) follows from the definition
of the spectral norm (and the fact that $\langle\langle\rho_{0}|\rho_{0}\rangle\rangle=1$),
and the fourth line is a consequence of the submultiplicativity of
$\left\Vert \ast\right\Vert $ and the triangle inequality. In the
last line we apply the unitary invariance of this norm. Writing $\sum_{k=2}^{\infty}\frac{1}{k!}\left(\intop_{0}^{T}dt\left\Vert \mathcal{L}(t)\right\Vert \right)^{k}$
as $e^{\intop_{0}^{T}dt\left\Vert \mathcal{L}(t)\right\Vert }-1-\intop_{0}^{T}dt\left\Vert \mathcal{L}(t)\right\Vert $,
we obtain Eq. (\ref{eq:41 error bound 1}).

Let us now derive the bound (\ref{eq:42 error bound 2}). We express
the difference $e^{\Omega(2T)}-e^{\chi}$ as $e^{\Omega(2T)}-e^{\chi}=\left(e^{\Omega(2T)}-I_{L}-\chi\right)-\sum_{k=2}^{\infty}\frac{1}{k!}\chi^{k}$,
and apply the triangle inequality to obtain 
\begin{equation}
\left\Vert e^{\Omega(2T)}-e^{\chi}\right\Vert \leq\left\Vert e^{\Omega(2T)}-I_{L}-\chi\right\Vert +\sum_{k=2}^{\infty}\frac{1}{k!}\left\Vert \chi\right\Vert ^{k}.\label{eq:S3.7}
\end{equation}
Now, keeping in mind that $\chi=\Omega_{1}(2T)$, the term $\left\Vert e^{\Omega(2T)}-I_{L}-\chi\right\Vert $
is analogous to $\left\Vert I_{L}-e^{\Omega(T)}+\Omega_{1}\right\Vert $
in the second line of Eq. (\ref{eq:S3.6}). Specifically, in both
cases we have the subtraction between the exact evolution and the
sum of the corresponding first Magnus term plus $I_{L}$. As a consequence,
the application of the Dyson expansion 
\begin{equation}
e^{\Omega(2T)}=\sum_{k=0}^{\infty}\frac{1}{k!}\intop_{0}^{2T}dt_{1}\intop_{0}^{2T}dt_{2}...\intop_{0}^{2T}dt_{k}\mathcal{T}\left(\mathcal{L}_{\textrm{ext}}^{\prime}\right)^{int}(t_{1})...\left(\mathcal{L}_{\textrm{ext}}^{\prime}\right)^{int}(t_{k}),\label{eq:S3.8}
\end{equation}
and the same steps followed in Eq. (\ref{eq:S3.6}), lead to
\begin{align}
\left\Vert e^{\Omega(2T)}-I_{L}-\chi\right\Vert  & \leq\sum_{k=2}^{\infty}\frac{1}{k!}\left(\intop_{0}^{2T}dt\left\Vert \mathcal{L}_{\textrm{ext}}^{\prime}(t)\right\Vert \right)^{k}\nonumber \\
 & =e^{\intop_{0}^{2T}dt\left\Vert \mathcal{L}_{\textrm{ext}}^{\prime}(t)\right\Vert }-1-\intop_{0}^{2T}dt\left\Vert \mathcal{L}_{\textrm{ext}}^{\prime}(t)\right\Vert .\label{eq:S3.9}
\end{align}

Since $\mathcal{L}_{\textrm{ext}}^{\prime}=\mathcal{L}_{\textrm{ext}}(t)-i\delta H_{L}(t)$
(cf. Eqs. (\ref{eq:40 extend dissipator}) an (\ref{eq:S1.10})),
Eq. (\ref{eq:S3.9}) yields half of the bound (\ref{eq:42 error bound 2}).
The other half results from the series $\sum_{k=2}^{\infty}\frac{1}{k!}\left\Vert \chi\right\Vert ^{k}$.
By expressing $\chi=\Omega_{1}(2T)$ as in Eq. (\ref{eq:S1.13}),
it follows that 
\begin{align}
\sum_{k=2}^{\infty}\frac{1}{k!}\left\Vert \chi\right\Vert ^{k} & =\sum_{k=2}^{\infty}\frac{1}{k!}\left\Vert \int_{0}^{2T}dt\left(\mathcal{L}_{\textrm{ext}}^{\prime}\right)^{int}(t)\right\Vert ^{k}\nonumber \\
 & \leq\sum_{k=2}^{\infty}\frac{1}{k!}\left(\intop_{0}^{2T}dt\left\Vert \mathcal{L}_{\textrm{ext}}^{\prime}(t)\right\Vert \right)^{k}\nonumber \\
 & =e^{\intop_{0}^{2T}dt\left\Vert \mathcal{L}_{\textrm{ext}}^{\prime}(t)\right\Vert }-1-\intop_{0}^{2T}dt\left\Vert \mathcal{L}_{\textrm{ext}}^{\prime}(t)\right\Vert .\label{eq:S3.10}
\end{align}
Therefore, $\left\Vert e^{\Omega(2T)}-e^{\chi}\right\Vert $ in (\ref{eq:S3.7})
is bounded by 
\begin{align}
\left\Vert e^{\Omega(2T)}-e^{\chi}\right\Vert  & \leq2\left(e^{\intop_{0}^{2T}dt\left\Vert \mathcal{L}_{\textrm{ext}}^{\prime}(t)\right\Vert }-\intop_{0}^{2T}dt\left\Vert \mathcal{L}_{\textrm{ext}}^{\prime}(t)\right\Vert -1\right)\nonumber \\
 & =2\left(e^{\intop_{0}^{2T}dt\left\Vert \mathcal{L}_{\textrm{ext}}(t)-i\delta H_{L}(t)\right\Vert }-\intop_{0}^{2T}dt\left\Vert \mathcal{L}_{\textrm{ext}}(t)-i\delta H_{L}(t)\right\Vert -1\right).\label{eq:S3.11}
\end{align}

\end{document}